\documentclass[letter]{aa} 

\usepackage{graphicx}
\usepackage{txfonts}
\usepackage{inputenc}
\usepackage{verbatim} 
\usepackage[normalem]{ulem} 

\begin{document} 

   \title{Linear polarization in the nucleus of M87 at 7\,mm and 1.3\,cm}
   \titlerunning{Linear polarization in the nucleus of M87}
   \author{E. Kravchenko \inst{1,2}
          \and M. Giroletti \inst{1}
          \and K. Hada \inst{3,4}
          \and D. L. Meier \inst{5,6}
          \and M. Nakamura \inst{7}
          \and J. Park\inst{7,8}
          \and R. C. Walker \inst{9}
          }
  \institute{INAF Istituto di Radioastronomia, Via P. Gobetti, 101, Bologna 40129, Italy\\
  \email{evgeniya.kravchenko@inaf.it}
  \and Astro Space Center, Lebedev Physical Institute, Profsouznaya 84/32, Moscow 117997, Russia
  \and Mizusawa VLBI Observatory, National Astronomical Observatory of Japan, 2-12 Hoshigaoka, Mizusawa, Oshu, Iwate 023-0861, Japan
  \and Department of Astronomical Science, The Graduate University for Advanced Studies (SOKENDAI), 2-21-1 Osawa, Mitaka, Tokyo 181-8588, Japan
  \and Department of Astronomy, Caltech, Pasadena, CA 91125, USA
  \and Jet Propulsion Laboratory, Pasadena, CA 91109, USA
  \and Institute of Astronomy and Astrophysics, Academia Sinica, P.O. Box 23-141, Taipei 10617, Taiwan
  \and Department of Physics and Astronomy, Seoul National University, Gwanak-gu, Seoul 08826, Republic of Korea
  \and National Radio Astronomy Observatory, Socorro, NM 87801, USA
  }

  \date{Received December 13, 2019; accepted April 24, 2020; published May 15, 2020}

  \abstract{We report on high angular resolution polarimetric observations of the nearby radio galaxy M87 using the Very Long Baseline Array at 24\,GHz ($\lambda=1.3$~cm) and 43\,GHz ($\lambda=7$~mm) in 2017--2018. New images of the linear polarization substructure in the nuclear region are presented, characterized by a two-component pattern of polarized intensity and smooth rotation of the polarization plane around the 43\,GHz core. From a comparison with an analogous dataset from 2007, we find that this global polarization pattern remains stable on a time interval of 11\,yr, while showing smaller month-scale variability. We discuss the possible Faraday rotation toward the M87 nucleus at centimeter to millimeter wavelengths. These results can be interpreted in a scenario where the observed polarimetric pattern is associated with the magnetic structure in the confining magnetohydrodynamic wind, which also serves as the source of the observed Faraday rotation.}
  
   \keywords{galaxies: active -- radio continuum: galaxies -- galaxies: individual M~87 -- galaxies: nuclei -- galaxies: jets -- galaxies: magnetic fields}
   \maketitle
%

\section{Introduction}

Relativistic jets are a common feature of extreme active galactic nuclei (AGNs).
General-relativistic, magnetohydrodynamic (GRMHD) simulations show that an accreting supermassive black hole (SMBH) and strong magnetic fields play dynamically important roles in their production  \citep[e.g.,][]{meier_etal01, 2006MNRAS.368.1561M, 2011MNRAS.418L..79T}.
Rotational energy from the central accreting black hole (BH) is likely electromagnetically extracted via the Blandford-Znajek (BZ) mechanism \citep{1977MNRAS.179..433B}, and is transferred away along the magnetic field lines connected to the BH event horizon.
This will lead to the formation of a narrow, relativistic, and highly magnetized jet. 
At the same time, field lines anchored to the inner regions of the accretion flow can launch broader outflows and winds through the Blandford--Payne (BP) process \citep{1982MNRAS.199..883B}.
The  hypothesis that the BZ and BP mechanisms act simultaneously in relativistic jets is supported by high-resolution studies of radio galaxies, such as Cygnus A \citep{2016AA...588L...9B} and 3C84 \citep {2018NatAs...2..472G}.

Observations of the giant elliptical galaxy M87 taken more than 100 years ago marked the first discovery of an extragalactic relativistic jet \citep{1918PLicO..13....9C}.
At a distance of $16.7\pm0.5$~Mpc \citep{2010AA...524A..71B} the radio galaxy hosts the nearest and largest active SMBH, with a mass of $M_\textrm{BH}=6.5\times10^9 M_\odot$ \footnote{This provides an angular-to-spatial scale ratio of $1~\mathrm{milli arcsecond ~(mas)} \approx 0.08 \mathrm{pc} \approx 130 ~R_{\rm S}$, where $R_\mathrm{S}=2GM/c^2 \equiv 2 r_{\rm g}$ is Schwarzschild radius and $r_{\rm g}$ is gravitational radius.} \citep{2019ApJ...875L...1E}, which makes M87 one of the most widely studied AGNs. 
Estimates of the jet production efficiency of $\geq110$\% \citep{2019ApJ...871..257P} and close-to-maximum magnetic flux threading the black hole \citep{2014Natur.510..126Z} indicate that the M87 jet is magnetically dominated near the base, implying that the MAD  \citep[Magnetically Arrested Disk,][]{2003PASJ...55L..69N, 2011MNRAS.418L..79T} scenario might be at work. 
Moreover, extensive VLBI studies show evidence of an acceleration of the flow taking place within $10^6~R_{\rm S}$ \citep{2014ApJ...781L...2A, 2016AA...595A..54M, 2019ApJ...887..147P}, which implies MHD conversion of the initial Poynting flux into bulk kinetic energy \citep{1992ApJ...394..459L}.

Polarimetric Very Long Baseline Interferometer (VLBI) observations are vitally important for tracing magnetic fields and testing theoretical models of black hole powered jets. 
From the high-dynamic-range VLA and HST images, strong polarization in excess of $\gtrsim60$\% has been found in the M87 jet progressively downstream from the nucleus ($\gg100$\,pc) at radio, optical, and UV wavelengths \citep{1988AA...202L..23S, 1989ApJ...340..698O,1997AA...317..637C, 1999AJ....117.2185P}, indicating the presence of a highly ordered magnetic field.
It has been suggested that a system of MHD shocks in a helical magnetic field can explain the formation of the conical jet in M87 at the kiloparsec scale \citep{2010ApJ...721.1783N,2014ApJ...785..152N}.
Meanwhile, from a marginal detection of  individual linearly polarized features, \citet{2016ApJ...817..131H} and \citet{2018ApJ...862..151H} reported a weak polarization (or high depolarization) in the nuclear jet region of M87 at the level of about 2-4\%.
 This may be due to the action of a dense foreground Faraday screen, which was revealed by high rotation measure (RM) in excess of $\sim10^3$\,rad~m$^{-2}$  \citep{2002ApJ...566L...9Z, 2019ApJ...871..257P}. 
There has been growing evidence for the existence of substantial nonrelativistic uncollimated gas outflow (wind) in M87, which is launched from the accretion flow \citep[e.g.,][]{2006MNRAS.368.1561M, 2012ApJ...761..129Y}.
This wind is probably a dominant source of the observed RM within the Bondi radius \citep{2019ApJ...871..257P}, and is a primary candidate for shaping, collimation, and acceleration of the jet in the MHD process \citep{2016ApJ...817..131H,2018ApJ...868..146N}.
In the most promising adiabatic inflow–outflow scenario \citep[ADIOS,][]{1999MNRAS.303L...1B,2004MNRAS.349...68B,2012MNRAS.420.2912B}, this wind is invoked to take away a significant amount of mass and energy from the disk, resulting in the low accretion rate observed in M87 \citep{2003ApJ...582..133D}.

\begin{table}
\caption{Details of VLBA observations and beam sizes for the images shown in Fig.~\ref{f:pol}.}
\label{tab:t1} 
\centering     
\begin{tabular}{llcc}
\hline\hline              
Project &  Date   & $\nu$ & Beam \\
code &  & [GHz] & [mas $\times$ mas, \degr{}] \\
\hline 
BW088A  & 2007--01--27 & 43.15 & 0.41 $\times$ 0.20, -10.7 \\
BW088G  & 2007--05--10 & 43.15 & 0.46 $\times$ 0.21, -16.6 \\
BG251A  & 2017--05--05 & 43.15 & 0.41 $\times$ 0.24,~~ 0.2   \\
BG250A$^{a}$  & 2018--04--28 & 43.15 & 0.39 $\times$ 0.21, ~-1.8 \\
BG250AK$^{b}$ & 2018--04--28 & 23.75 & 0.76 $\times$ 0.42, ~-9.8 \\
BG250B1$^{c}$ & 2018--05--25 & 43.15 & 0.38 $\times$ 0.17, ~-9.6 \\
\hline
\end{tabular}
\tablefoot{The following antennas are missed in observations: $^{a}$ - Hancock and Pie Town; $^{b}$ - Pie Town; $^{c}$ - Owens Valley.}
\end{table}

Recently, \citet{2018ApJ...855..128W} presented high-resolution polarimetric images of the M87 jet base, using observations performed with the upgraded Very Long Baseline Array (VLBA) at 43\,GHz in 2007.
These were the first spatially resolved images of the linearly polarized intensity in the nucleus of M87, which revealed a complex polarized substructure at a resolution of $\sim20~R_{\rm S}$.
In this Letter we extend the dataset of \citet{2018ApJ...855..128W} with new observations which were carried out using the VLBA at 24 and 43\,GHz in 2017 and 2018 toward the nucleus of M87.

\section{Observations and data reduction}
\label{sec:obs}
We summarize the details of the observations in Table~\ref{tab:t1}.
VLBA experiments were performed during five epochs, in 2007, 2017, and 2018, quasi-simultaneously at 24 and 43\,GHz.
A more detailed description of the 2007 dataset, which was obtained before the upgrade of VLBA hardware using a recording rate of 256 Mbps, is given by \citet{2018ApJ...855..128W}.
The observations in 2017 and 2018 were recorded at a rate of 2048\,Mbps in dual-polarization mode (right and left circular polarizations, RCP and LCP), and correlated with the VLBA software correlator in Socorro \citep{2011PASP..123..275D}.
The full bandwidth (256\,MHz per polarization) was split into $8\times32$\,MHz intermediate frequency (IF) channels.
The total on-source time for the target was about 1.7 hours at 24\,GHz, and 6 hours at 43\,GHz.
The initial data reduction was performed in the NRAO Astronomical Image Processing System \citep[AIPS,][]{aips}, following the standard procedures of VLBI data reduction \citep{VLBAmemo36,VLBAmemo37}.
Deconvolution and self-calibration algorithms, implemented in Difmap \citep{1997ASPC..125...77S}, were used for phase and amplitude calibration and for constructing the final images.
We adopted a typical amplitude calibration error of 10\% for both frequencies.
Details on the instrumental polarization calibration are summarized in Appendix~\ref{app:calib}.

\section{Results}
\label{sec:res}
\subsection{Linear polarization structure in the nucleus}
\label{s:lpol}
\begin{figure*}
   \centering
   \includegraphics[width=0.54\columnwidth,angle=-90]{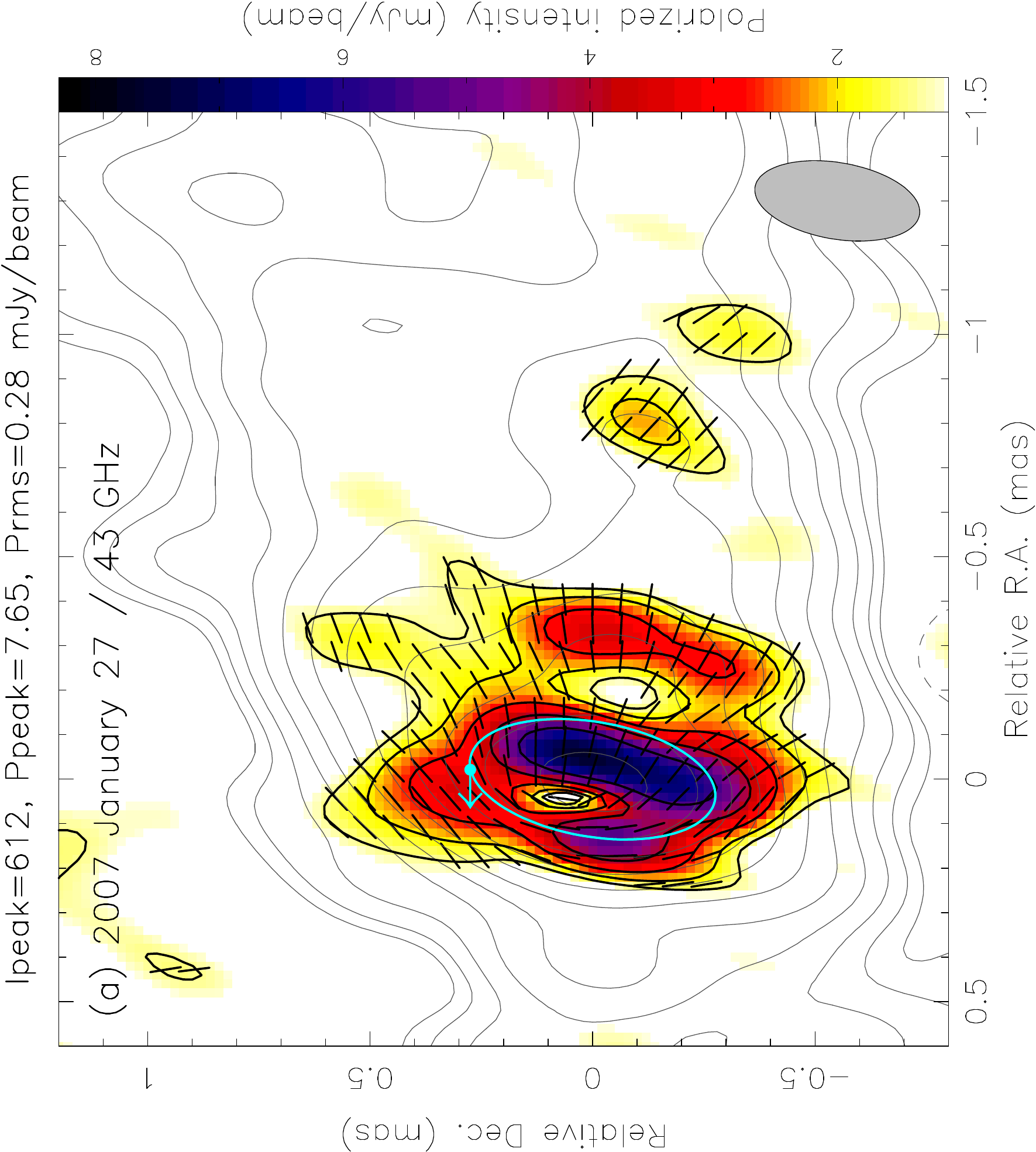}\quad
   \includegraphics[width=0.54\columnwidth,angle=-90]{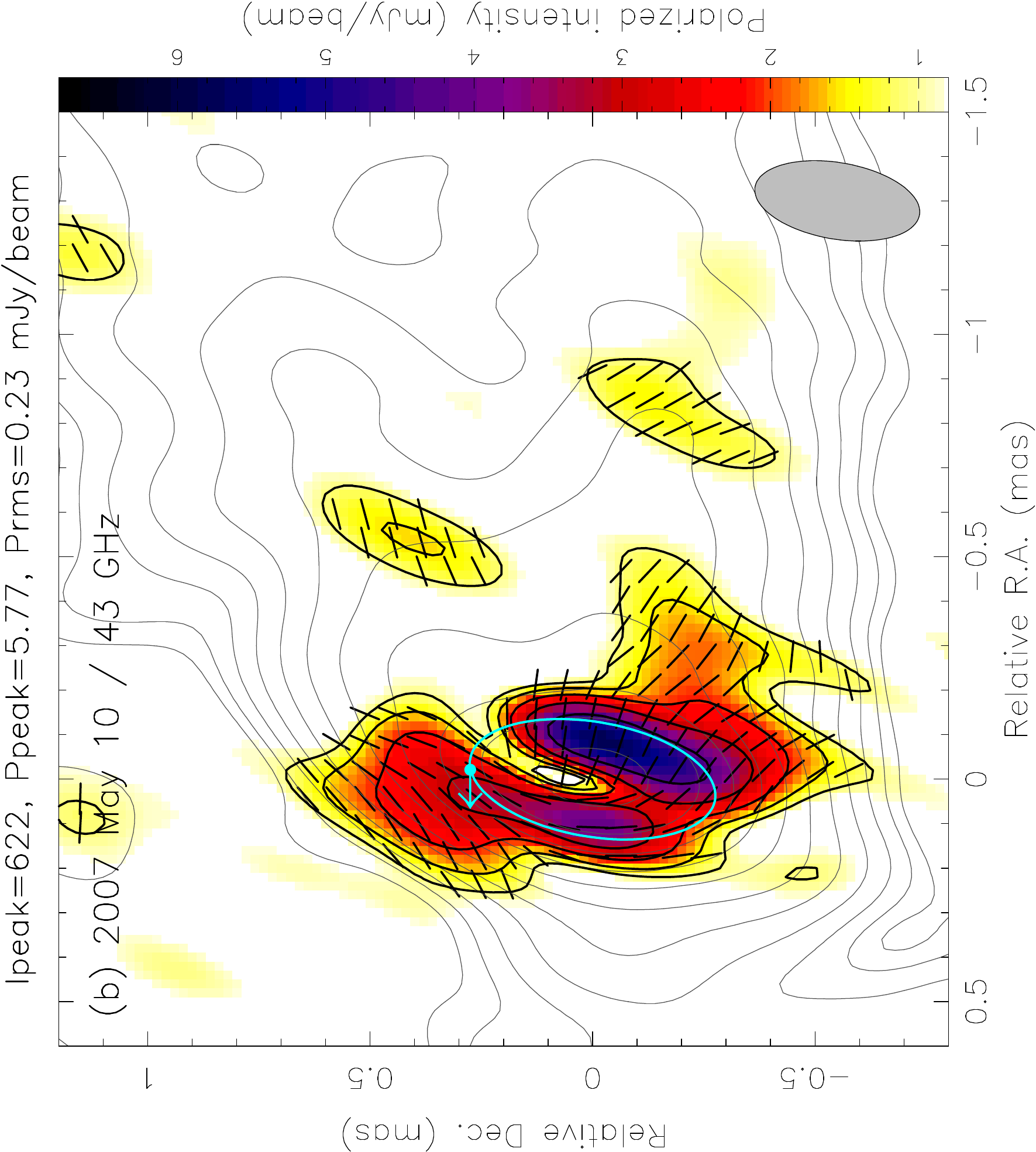}\quad
   \includegraphics[width=0.54\columnwidth,angle=-90]{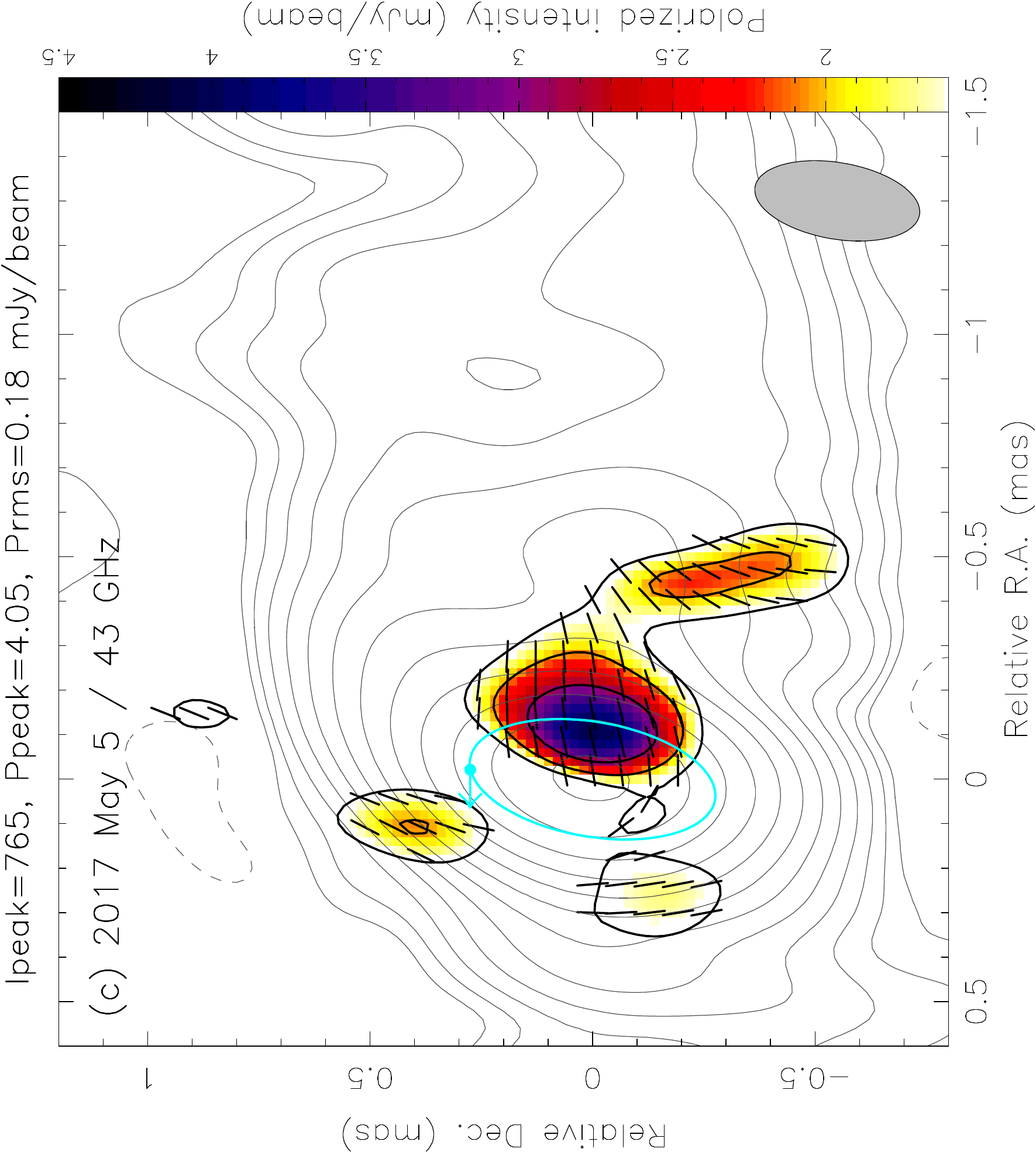}\\
   \vspace{7px}
   \includegraphics[width=0.54\columnwidth,angle=-90]{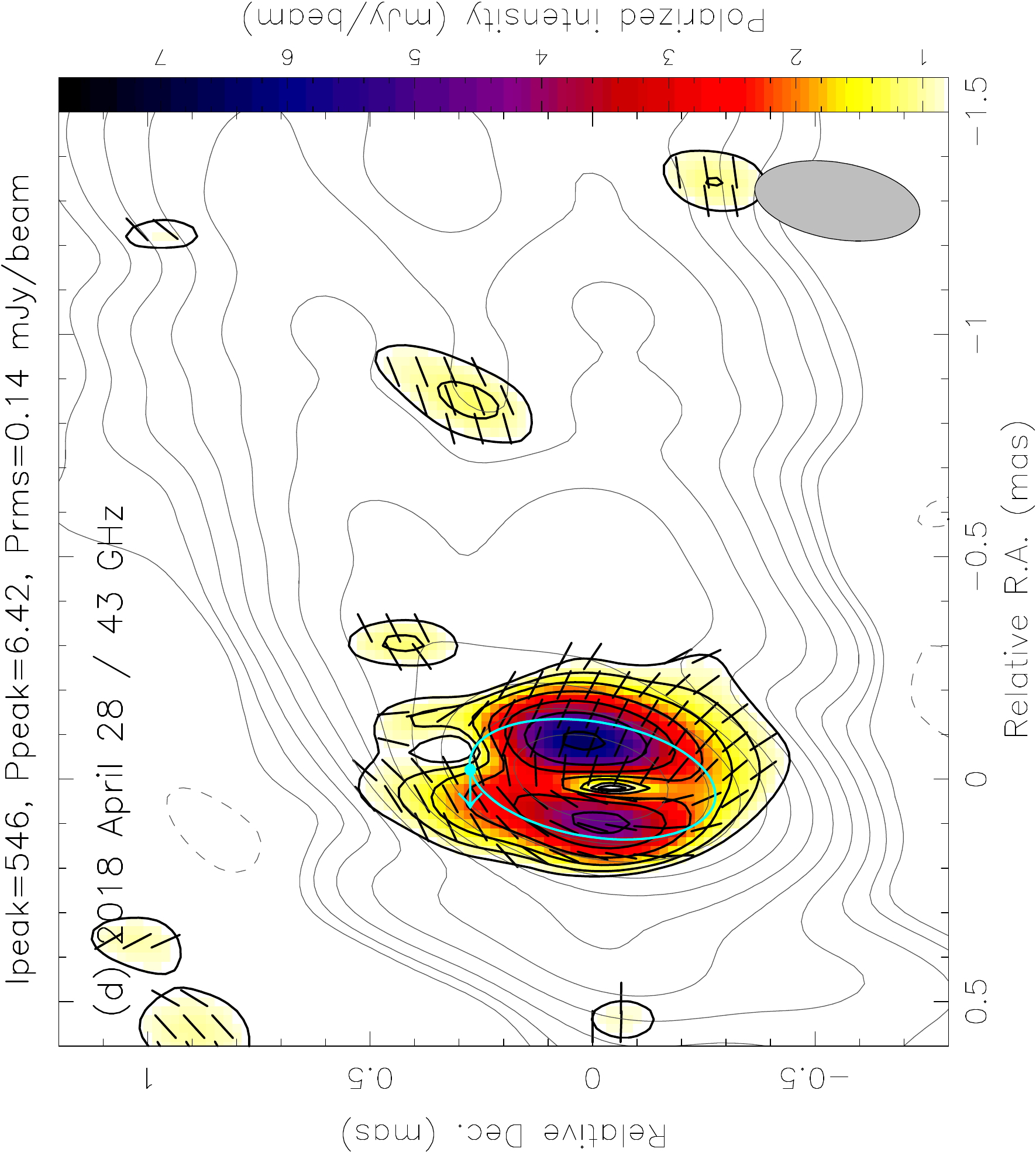}\quad
   \includegraphics[width=0.54\columnwidth,angle=-90]{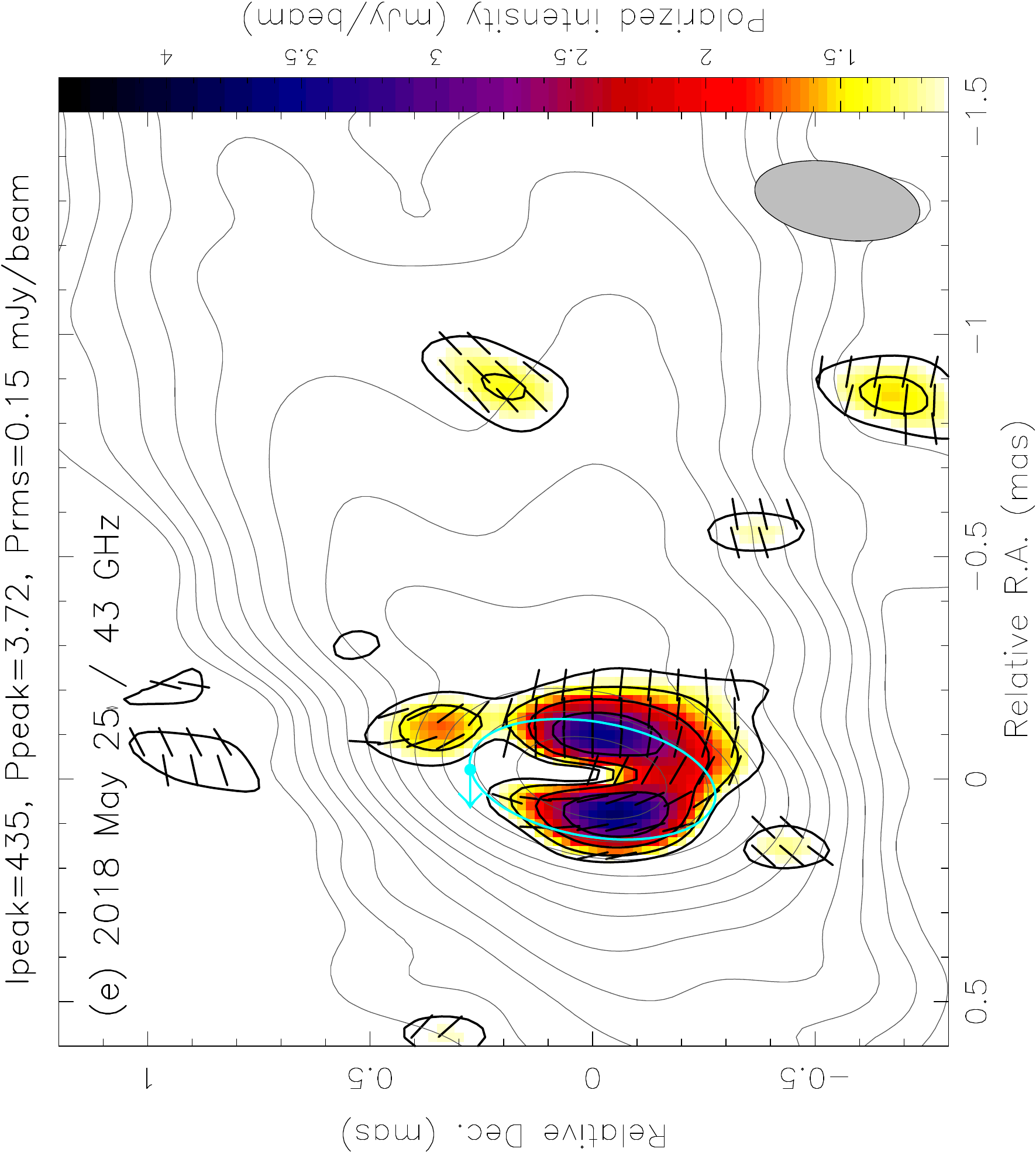}\quad
   \includegraphics[width=0.54\columnwidth,angle=-90]{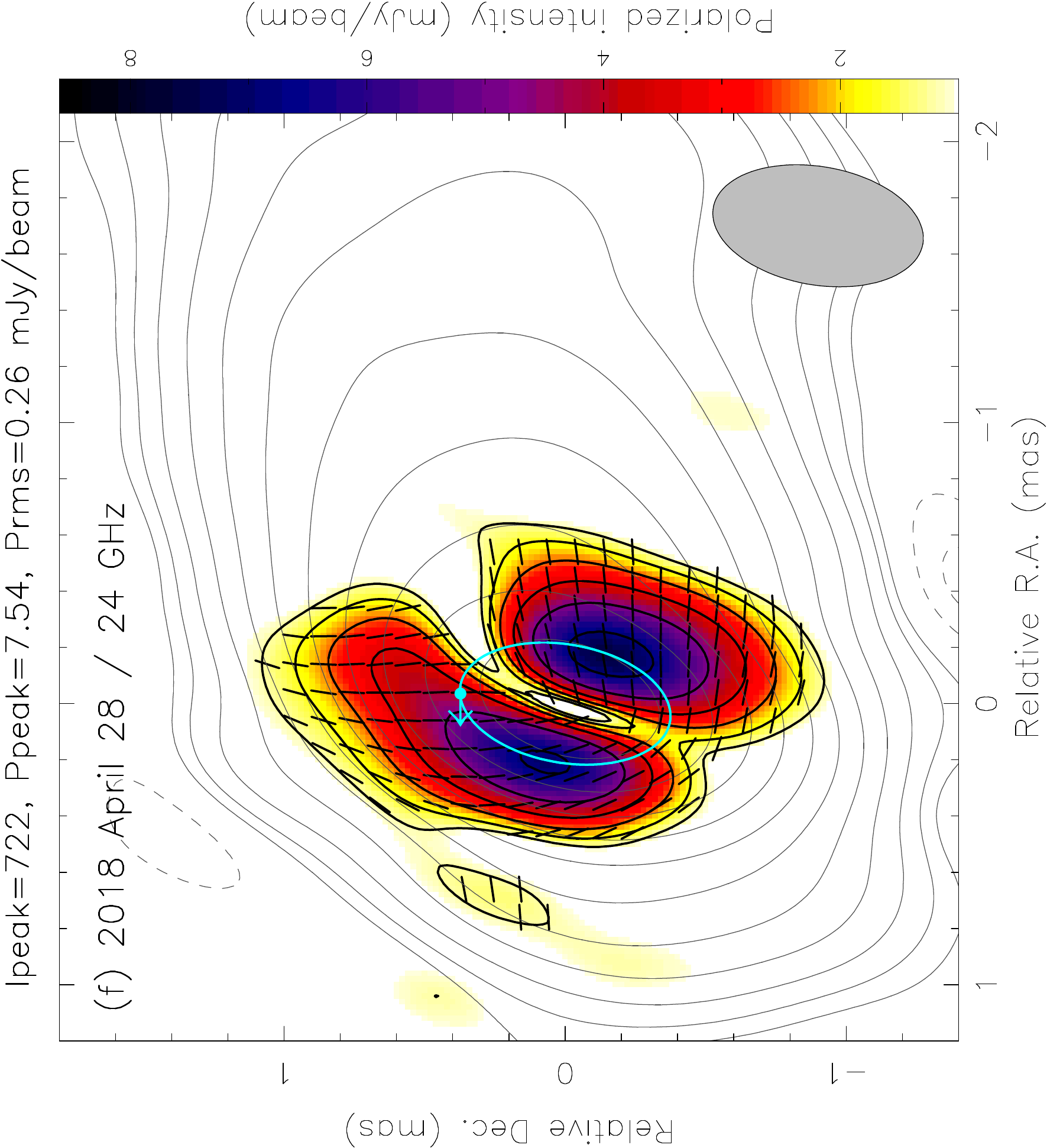}\\
   \caption{VLBA 24\,GHz (f) and 43\,GHz (a-e) polarimetric images of M87 in 2007--2018. The color map and black contours indicate the observed polarized intensity, vectors show the observed orientation of EVPA (uncorrected for Faraday RM), and gray contours denote the total intensity distribution. The synthesized beam is 0.76$\times$0.42\,mas at $-$10\degr{} at 24\,GHz and 0.38$\times$0.17\,mas at $-10$\degr{} at 43\,GHz, and is illustrated by the shaded ellipse. Stokes $I$ contours increase by factors of 2, starting from 3.5\,$\sigma_{\rm rms}$, where $\sigma_{\rm rms}$ is 134 (a), 133 (b), 94 (c), 89 (d), 74 (e), and 128 (f) $\mu$Jy/beam. $P$ contours start at 1.37 (a), 1.16 (b), 1.45 (c), 0.75 (d), 0.96 (e), and 1.17 (f) mJy/beam, and increase by factors of $\sqrt{2}$. The blue ellipse specifies the region for which distributions (Figure~\ref{f:slice} and Figure~\ref{f:frm}) have been computed, and the arrow indicates the start point and direction used for this analysis.}
   \label{f:pol}
\end{figure*}

The resulting linearly polarized flux density maps at 24 and 43\,GHz of the core region of the M87 jet are shown in Fig.~\ref{f:pol}.
For the analysis, all 43\,GHz images were restored with the same beam size of the BG250B1 dataset (see Table~\ref{tab:t1}) and aligned at the position of the peak in total intensity. 
The resultant polarization structure is localized within $\thicksim0.7$ (43\,GHz) -- 1 (24\,GHz)\,mas around the core region, and is consistent through all epochs of observations. 
It is characterized by two different polarized features, located upstream and downstream of the peak in total intensity, and by a minimum of the linear polarization $P$ in between.
The average degree of polarization is about 2-3\% (see Fig.~\ref{f:mdegr}). 
The electric vector position angle (EVPA) in the downstream component is oriented along the approaching jet direction, and then gradually rotates around the peak in total intensity, with the orientation of the electric vectors changing by about 90\degr{}.
The maximum value of the linear polarization is observed at a level of $7.4\pm0.8$\,mJy beam$^{-1}$.

For better visualization of the polarization structure, in Fig.~\ref{f:slice} we show the observed EVPAs (i.e., uncorrected for Faraday rotation) and linearly polarized flux density as a function of the position angle (PA) of the pixels with respect to the centroid of the core.
The centroid is illustrated in Fig.~\ref{f:pol}, and the EVPAs and $P$ are measured counterclockwise starting from the north. 
Smooth rotation of the linear polarization vectors, reaching almost 180\degr{} (when assuming continuous rotation) can be seen. 
This global rotating trend is consistent throughout all observational epochs, and was apparently present for a time period of more than 11 yr.
This suggests the presence of a persistent magnetic field structure in this region.
The peak of linearly polarized intensity is localized at the approaching side of the jet, at approximately  0.1\,mas from the core for all epochs. 
Most likely, this emission region can be associated with the only polarized feature detected on the 86\,GHz VLBI polarimetric image \citep{2016ApJ...817..131H} at the same distance from the radio core ($P_{\rm peak} \sim 5.0$\,mJy\,beam$^{-1}$).

\subsection{Faraday rotation}
\label{s:frm}
Magnetized plasma located along the line of sight to the observer causes Faraday rotation \citep{burn66}, produced by a different propagation velocity of the left and right circularly polarized waves.
This results in a rotation of the intrinsic polarized plane ($\phi_{\rm int}$) by an angle equal to the square of the wavelength ($\lambda$) multiplied by the rotation measure (RM), which can be estimated from the observed dependence of the EVPA ($\phi$) on $\lambda$, such that
\begin{equation}
\phi = \phi_{\rm int} + RM\lambda^2.
\end{equation}

To obtain the RM image, we first convolved the Stokes $I$, $Q$, and $U$ images taken at 24 and 43\,GHz with the 24\,GHz 2018-04-28 beam size, and then aligned them at the position of optically thin emission regions (see Appendix~\ref{app:spi} for details).
The $n\pi$-ambiguity in the polarization vector was first solved under the assumption of the smallest Faraday rotation, which relies only on the analysis of data at two frequencies. 
The resultant average RM value amounts to $\thicksim 2\times10^3$\,rad\,m$^{-2}$ (RM image is given in Appendix~\ref{app:frm}).

Subsequently, we associated the linearly polarized feature of \citet{2016ApJ...817..131H}, which is located at $\thicksim0.1$\, mas downstream from the 86\,GHz core, with the same region in our dataset. 
This is reasonable, as the polarized emission peaks at this region in all our observations, and it shows only moderate variability over a long period of time (further details and justification of this approach are discussed in Appendix~\ref{app:pir}).
We find a $-$180\degr{} rotation of the polarization angle at 24\,GHz (see Fig.~\ref{f:frm} and Appendix~\ref{app:frm}), and overall better linear alignment of EVPA versus $\lambda^2$ in the 24 -- 230\,GHz range, consistent with the scenario of external Faraday rotation \citep{burn66, sokoloff_etal98}.

Using this approach, we estimate the rotation measure $RM \thicksim -(2.3\pm0.2)\times10^4$\,rad\,m$^{-2}$ at a position 0.1\,mas from the core.
This still represents the lower limit, due to n$\pi$-ambiguity and differences in time between 24, 43 and 86\,GHz observations.
Moreover, due to the complex structure, the accretion flow from the BH can  simultaneously be the source of synchrotron radiation and the Faraday screen \citep[e.g.,][]{2017Galax...5...54M}, implying strong depolarization and internal Faraday rotation, which will lead to nonlinear dependence of $\phi(\lambda^2)$.
However, this is difficult to address in our dataset due to sparse $\lambda^2$ coverage.

\section{Discussion}
\label{sec:disc}

\subsection{Jet sheath as a source of Faraday rotation}
\label{s:dfrm}

A magnetized screen (or sheath), found in close proximity and surrounding the jet, has been suggested as a plausible source of Faraday rotation in AGN jets \citep[e.g.,][]{2004ApJ...612..749Z, 2017MNRAS.467...83K}. 
GRMHD simulations show that an accreting BH system can naturally form a spine-sheath structure with a fast sheath that surrounds a slow spine as a consequence of mass accretion and BH physics \citep[e.g.,][]{2006MNRAS.368.1561M, 2013MNRAS.436.3741P, 2018ApJ...868..146N}.
In this scenario, differential rotation of the inner parts of the accretion disk or BH ergosphere will produce helical magnetic fields \citep{meier_etal01}. 
Due to systematic changes in the line-of-sight component of helical magnetic fields, their signatures can be detected through transverse RM gradients, which are indeed observed in AGN jets \citep[e.g.,][]{2002PASJ...54L..39A}. 
A number of observational effects, including apparent limb brightening \citep{2007ApJ...668L..27K} and kinematic profiles \citep{2016AA...595A..54M}, favor a two-component model for the M87 jet.

\begin{figure}
   \centering
   \includegraphics[width=0.56\columnwidth,angle=-90]{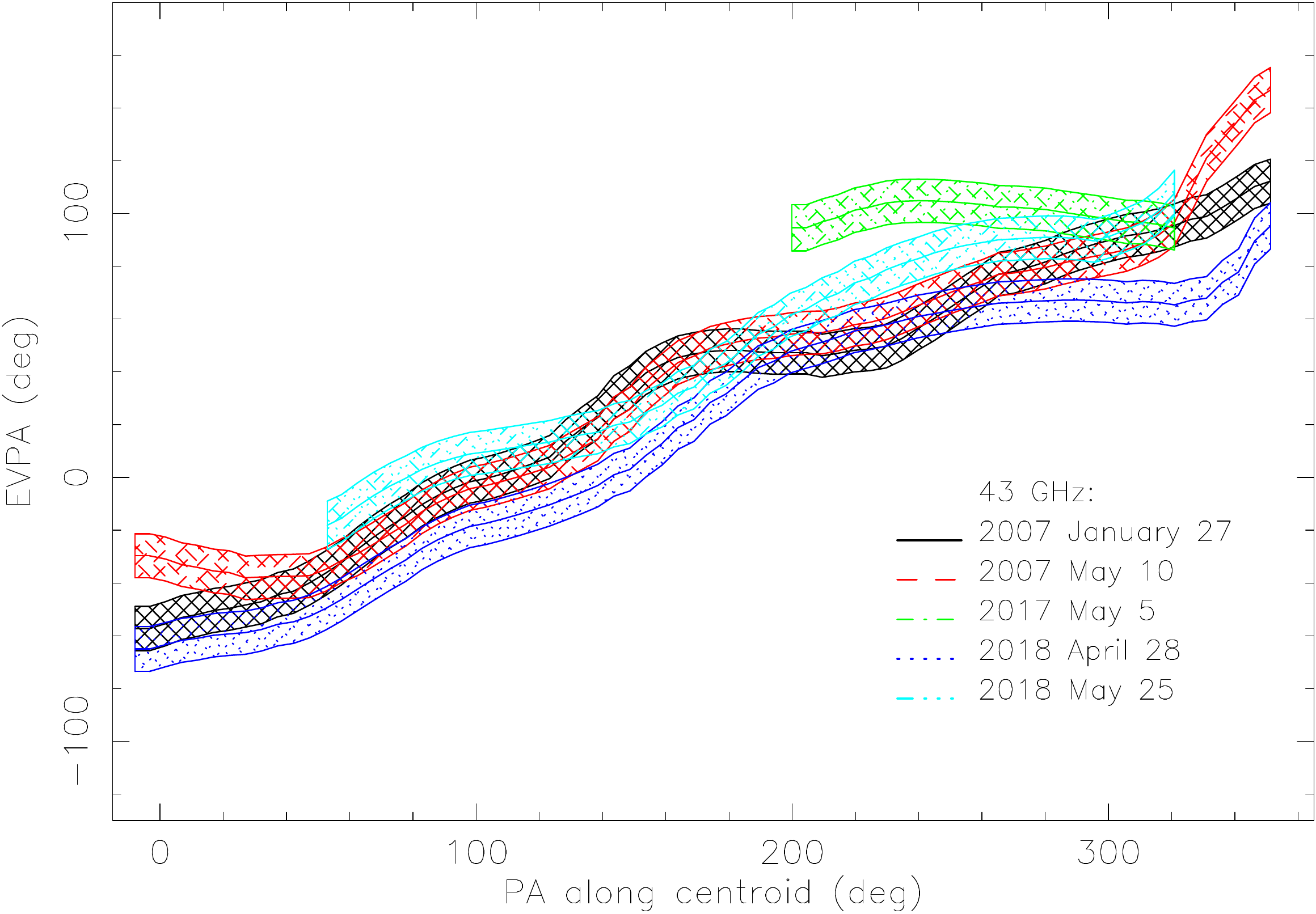}\\
   \includegraphics[width=0.56\columnwidth,angle=-90]{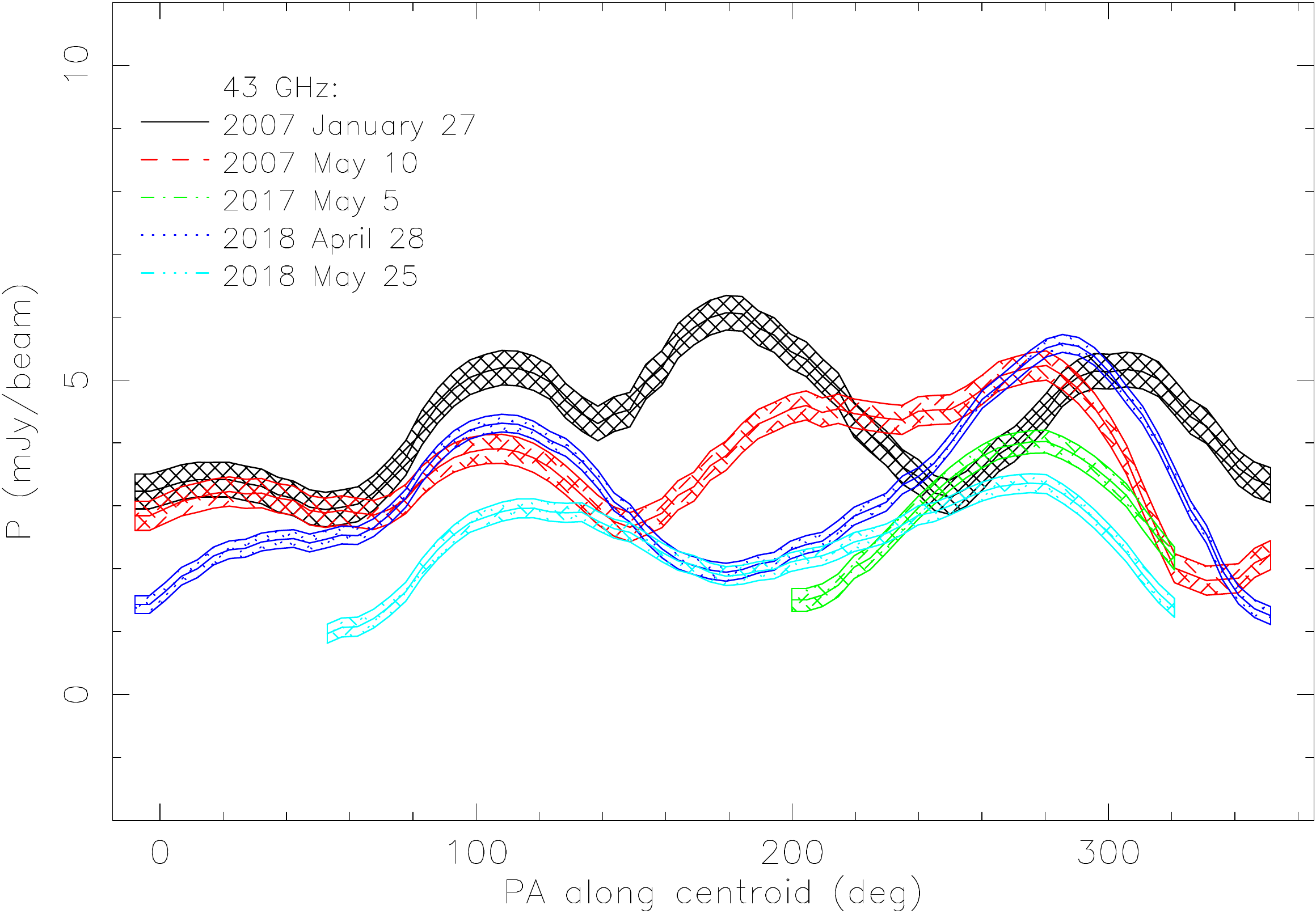}
   \caption{Observed EVPA ({\it top}) and linearly polarized flux density ({\it bottom}) at 43\,GHz along the ellipse indicated in Fig.~\ref{f:pol}. Hatched areas denote $1\sigma_{\rm rms}$ error.}
   \label{f:slice}
\end{figure}

In this case, the sheath will follow the same parabolic geometry as that observed for the jet, i.e., $z\propto r^{1.73}$, where $r$ is the radius of the jet emission and $z$ is the axial distance from the core \citep{2012ApJ...745L..28A, 2013ApJ...775...70H}.
Considering the decreasing gradient in electron density of the screen $\rho(r)\propto r^{-2} \propto z^{-1,16}$ and the absence of a velocity gradient across the jet, $RM \propto \int \rho(r) Bdl$.
For a toroidal magnetic field ($B\propto r^{-1}\propto z^{-0.58}$) this implies $RM \propto z^{-1.2}$.
If we assume dominance of the poloidal magnetic fields ($B\propto r^{-2}\propto z^{-1.16}$), which are  expected in the MAD-type accretion flow \citep{2011MNRAS.418L..79T}, we find $RM \propto z^{-1.7}$.
Extrapolating the RM value measured at 10\,mas downstream from the core \citep{2019ApJ...871..257P} to a position of 0.1\,mas, we obtain $|RM| \thicksim 3 \times 10^7$~rad~m$^{-2}$ and $|RM| \thicksim 2 \times 10^6$~rad~m$^{-2}$, for the poloidally and toroidally dominated magnetic fields, respectively. 
No sign reversals or transverse RM gradients have been observed in the M87 jet so far \citep{2002ApJ...566L...9Z, 2019ApJ...871..257P}, making invoking the jet sheath as the source of Faraday rotation problematic.

\subsection{Hot accretion flow}

In an alternative scenario, the Faraday rotation could be attributed to the hot accretion flow \citep{1976ApJ...204..187S}, which has been invoked to explain the low accretion state in M87.
The classical theory of an advection-dominated accretion flow \citep[ADAF, ][]{1977ApJ...214..840I, 1994ApJ...428L..13N} predicts a density profile of $\rho(r)\propto r^{-p}$, where $p=1.5$.
Numerical simulations suggest an even shallower density profile for the hot accretion flow, with $p=0.5-1$ \citep{2012ApJ...761..130Y, 2012ApJ...761..129Y}.
This is in good agreement with what was determined from the observations of low-luminosity AGNs within the Bondi radius, e.g., M87 itself \citep[$\rho\propto r^{-1}$,][]{2015MNRAS.451..588R}, Sgr~A* \citep[$\rho\propto r^{-1}$,][]{2013Sci...341..981W, 2013Natur.501..391E}, and NGC~3115 \citep[$\rho\propto r^{-1.03}$,][]{2012ApJ...761..129Y}.
These results suggest that only a small fraction of the falling material reaches the SMBH, while the rest is carried away by substantial winds, as predicted by the ADIOS model ($0.5<p<1.5$).

Analyzing VLBA polarimetric observations of M87 at 2--8\,GHz, \citet{2019ApJ...871..257P} found that the RM decrease with distance from the nucleus, within $5\times10^3-2\times10^5~R_{\rm S}$, is in good agreement with the gas density profile $\rho \propto r^{-1\pm0.11}$ when assuming spherical geometry for the flow.
This therefore implies $|RM|\sim10^6$~rad\,m$^{-2}$ within the central 0.1\,mas in the jet.
Meanwhile, the accretion flow may become significant only beyond some distance from the BH \citep[e.g., $\thicksim30r_{\rm g}$ for non-rotating BH;][]{2012MNRAS.426.3241N, 2014ARA&A..52..529Y}, leading to a departure from the $\rho\propto r^{-1}$ dependence. This would explain our low observed RM value.

\subsection{Magnetic field}

The EVPA orientation intrinsic to the source is difficult to determine, because of the relatively uncertain Faraday rotation measure. The value of $-2.3\times10^4$\,rad\,m$^{-2}$ found above will cause the rotation of the polarization plane by $\sim 55$\degr{} at 43\,GHz. Given that a) there is no 90\degr{} flip in EVPA (which would be expected due to an opacity change; see Appendix~\ref{app:opac} for details), b) the orientation of the observed polarization and associated magnetic field are perpendicular, and c) the magnetized plasma is nonrelativistic \citep{2019ApJ...887..147P}, the EVPAs in Fig.~\ref{f:pol} should therefore be rotated by $\sim35$\degr{} counterclockwise to highlight the magnetic field configuration. At the same time, the EVPAs that are wrapped around the core suggest a toroidal jet magnetic field geometry.

The image of the BH shadow in M87 obtained from the EHT observations at 230\,GHz \citep{2019ApJ...875L...4E} is consistent with the models of jet launching driven through the BZ process \citep{2019ApJ...875L...5E}. This therefore requires a magnetic field of at least $10^3$\,G threading the BH interiors \citep{2019ARAA..57..467B} in order to power the jet of higher $10^{43}$\,erg\,s$^{-1}$ or above \citep{2006MNRAS.370..981S, 2000ApJ...543..611O, 2012A&A...547A..56D}.
In a flux-conserving situation, poloidal and toroidal magnetic field components in the wind will decrease from near the rotation axis as $B_{\rm p}\propto r^{-2}$ and $B_{\rm t}\propto r^{-1}$, respectively.
Meanwhile, they should be of roughly   equal strength on the Alfven surface at the edge of the jet funnel (i.e., $10~r_{\rm g}$), therefore implying $B_{\rm p}\sim B_{\rm t} \approx 10^3 ~{\rm G} ~ (1 r_{\rm g} / 10 r_{\rm g})^2 \sim 10$\,G.
Given the deprojected distance of the 43\,GHz core from the black hole of $\sim35~r_{\rm g}$ \citep{2011Natur.477..185H}, when assuming a jet viewing angle of 18\degr{} \citep{2016AA...595A..54M}, a confining wind at this distance should therefore have a toroidal field strength of $B_{\rm t} \sim 10~{\rm G} ~(10 r_{\rm g} / 35 r_{\rm g}) \sim 3$\,G.
This is about the same order as the magnetic field at the same distance in an ADAF disk around a $\sim6\times10^9 M_\odot$ black hole accreting at $10^{-3}$ of the Eddington rate \citep{2012bhae.book.....M}. 
Therefore, our VLBA observations are consistent with the model of a confining MHD wind, and in agreement with the highly magnetically dominated state of the radio core, as observed at 43\,GHz \citep[$1~G\leq B_{\rm 43~GHz, core}\leq15~G$,][]{2014ApJ...786....5K}, 86\,GHz \citep[$B_{\rm 86~GHz, core}\sim8$G,][]{2016ApJ...817..131H}, and 230\,GHz \citep[$50~G\leq B_{\rm 230~GHz, core}\leq124~G$,][]{2015ApJ...803...30K}.

The thermal component of a confining disc wind determined by the accretion timescale of the ADAF and contributing to the RM can explain the observed persistent polarization profile well. Meanwhile, emission produced close to the BH or in the inner jet could give rise to more rapid flaring emission, which has been detected at the jet base of M87 several times before \citep[e.g.,][]{2009Sci...325..444A, 2012ApJ...746..151A, 2014ApJ...788..165H}.
This can explain the observed second-order, month-scale variability of polarized flux density seen in Fig.~\ref{f:slice}.

\begin{figure}
   \centering
   \includegraphics[width=0.55\columnwidth,angle=-90]{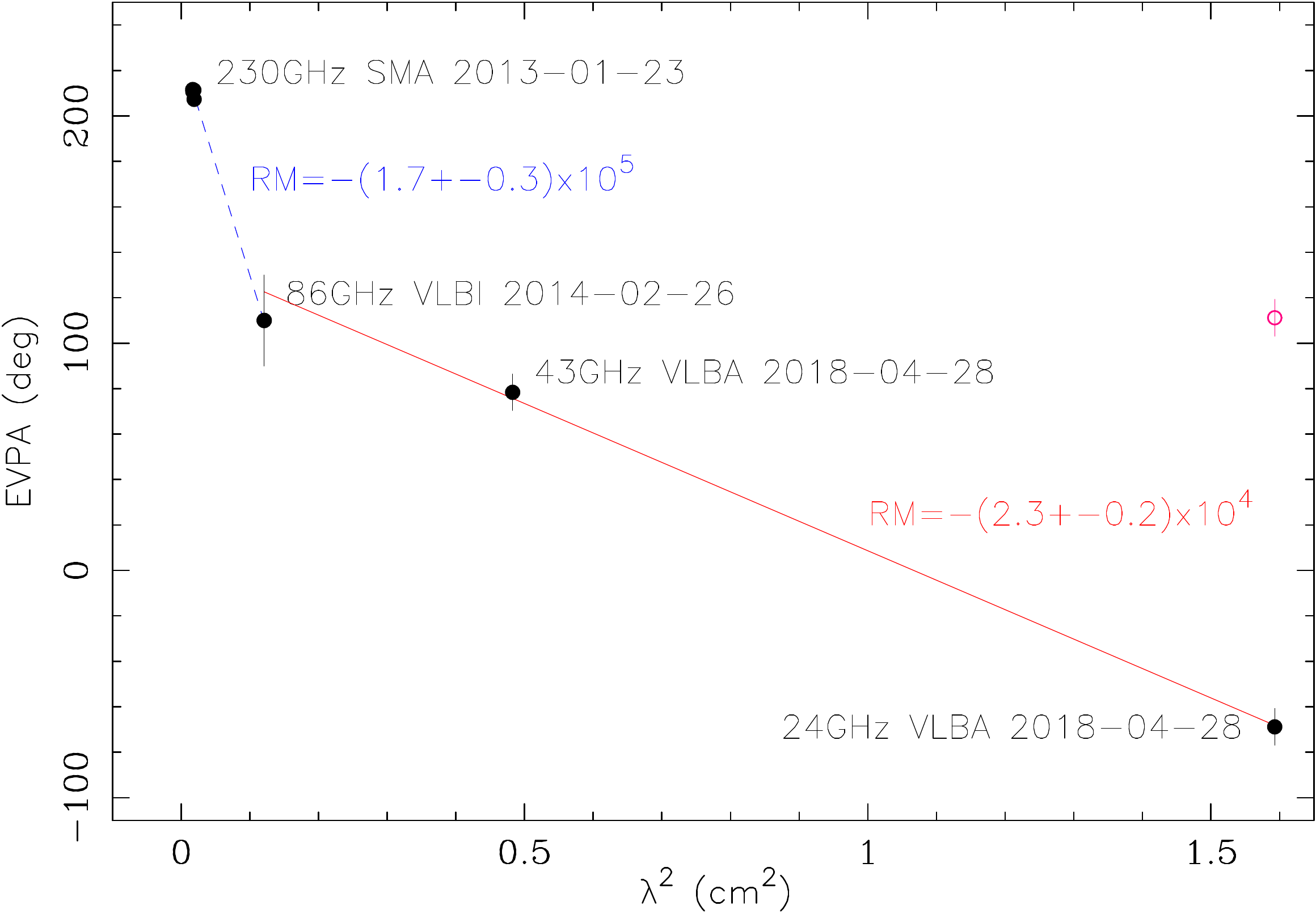}
   \caption{
   EVPA vs. $\lambda^2$ and RM fit at the position of 0.1\,mas downstream from the M87 radio core. The observed EVPA at 24\,GHz is shown by the empty red circle, while the value rotated by $-$180\degr{}, for a better RM fit, is shown by the filled black circle at the same value of $\lambda^2$. }
   \label{f:frm}
\end{figure}

\section{Conclusions}
\label{s:concl}
We present new polarimetric VLBI images of the M87 jet with spatially resolved, linearly polarized substructure in the core region.
The trend in rotation of electric vectors around the core centroid remains remarkably persistent throughout the time interval of 11\,yr.
At the same time, the linearly polarized intensity undergoes smaller month-scale variations.
Given that the polarized structure in the vicinity of the M87 nucleus remains moderately stable on a year-scale interval, the observed Faraday rotation in the 24 -- 86\,GHz range can be attributed to a magnetized screen, which is external to the emitting jet regions.
The observed polarization structure and Faraday rotation suggest that we are probing the magnetic structure in the adiabatic MHD wind.
Future EHT polarimetric observations, as well as actively ongoing detailed theoretical modeling, will soon shed light on the magnetic field properties in the vicinity of the M87 super massive black hole.
This will allow us to perform a detailed comparison of synthetic 43 and 230\,GHz images and to further test the jet models.

\begin{acknowledgements}
We thank the anonymous referee and Svetlana Jorstad for comprehensive comments and suggestions that helped to improve the manuscript.
This study makes use of 43 GHz VLBA data from the VLBA-BU Blazar Monitoring Program (VLBA-BU-BLAZAR; http://www.bu.edu/blazars/VLBAproject.html), funded by NASA through the Fermi Guest Investigator Program. 
The VLBA is an instrument of the National Radio Astronomy Observatory. 
The National Radio Astronomy Observatory is a facility of the National Science Foundation operated by Associated Universities, Inc. 
This work made use of the Swinburne University of Technology software correlator, developed as part of the Australian Major National Research Facilities Programme and operated under licence.
\end{acknowledgements}

\bibliographystyle{aa}
\bibliography{m87pol}

\begin{appendix}

\section{Calibration of instrumental polarization}
\label{app:calib}

\begin{figure*}
   \centering
   \includegraphics[width=0.46\columnwidth,angle=-90]{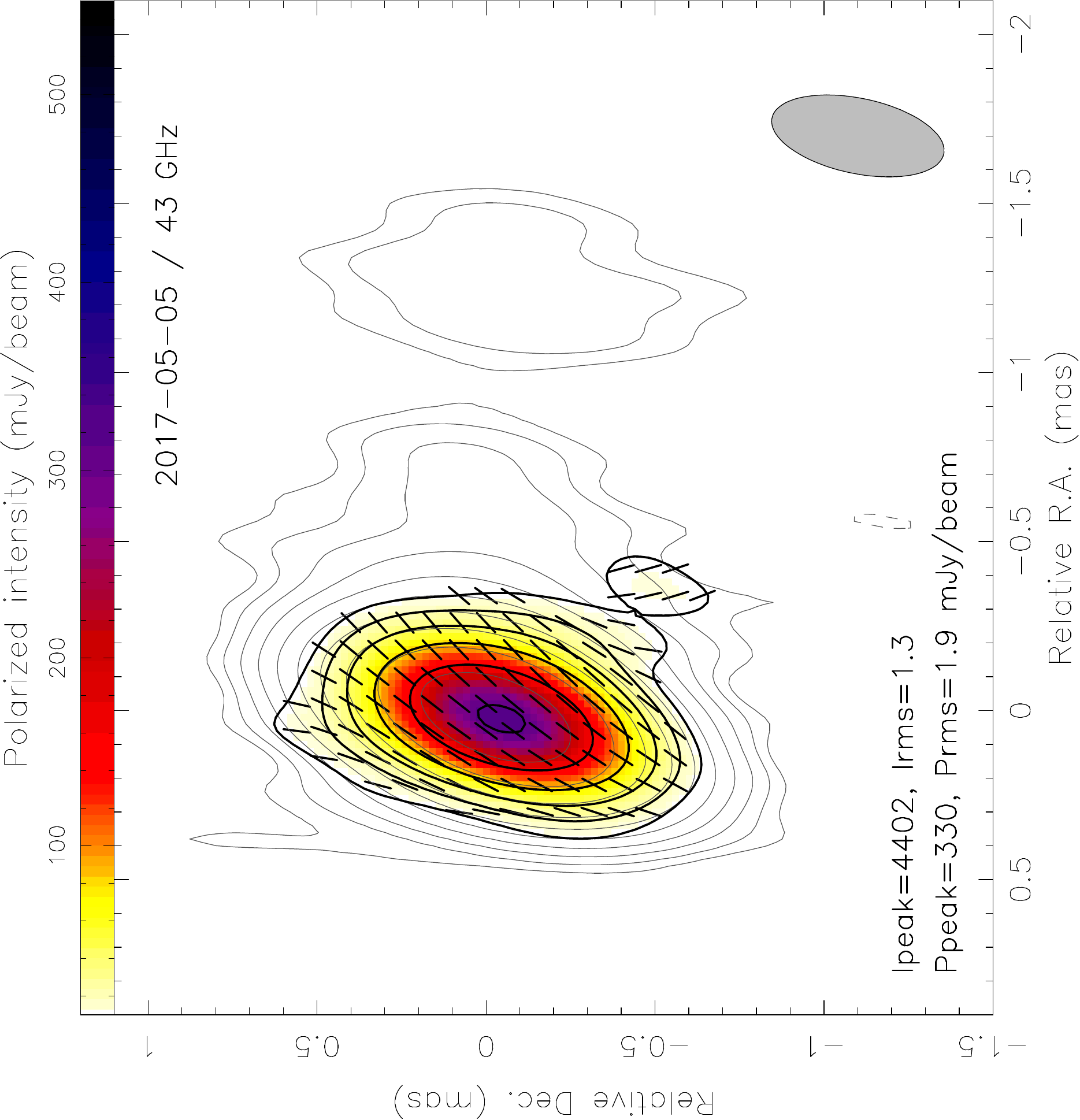}
   \includegraphics[width=0.46\columnwidth,angle=-90]{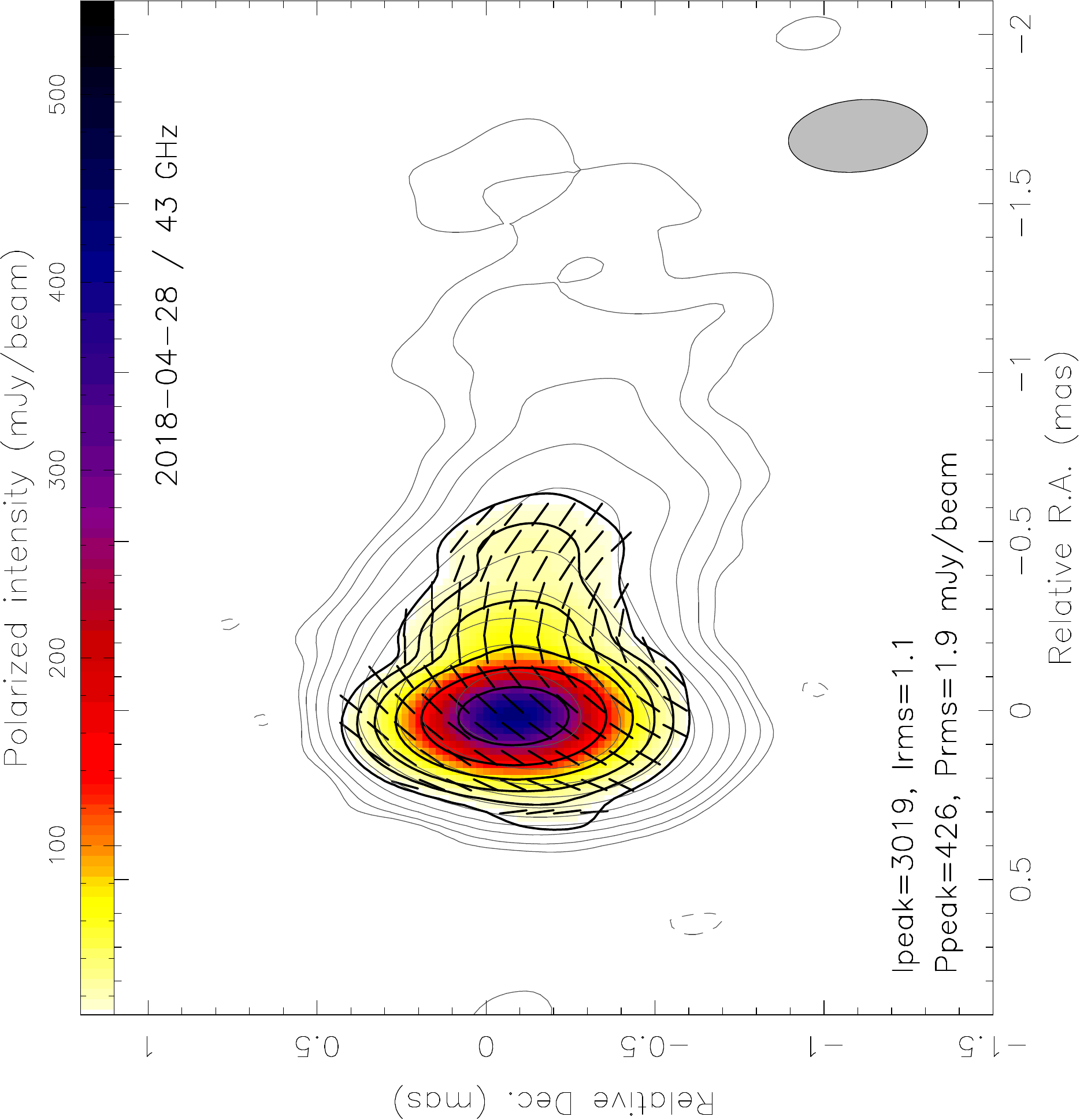}
   \includegraphics[width=0.46\columnwidth,angle=-90]{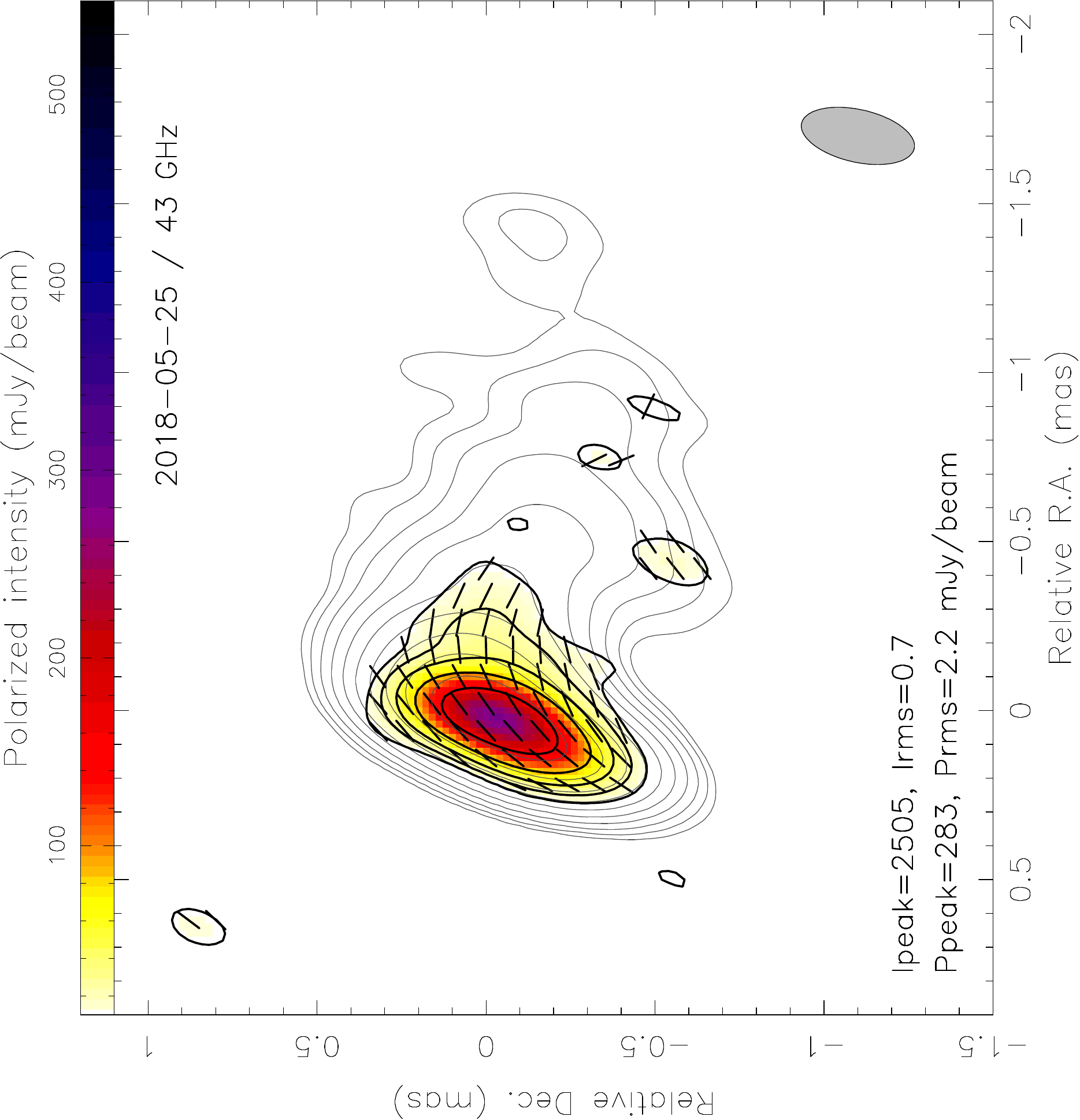}
   \includegraphics[width=0.46\columnwidth,angle=-90]{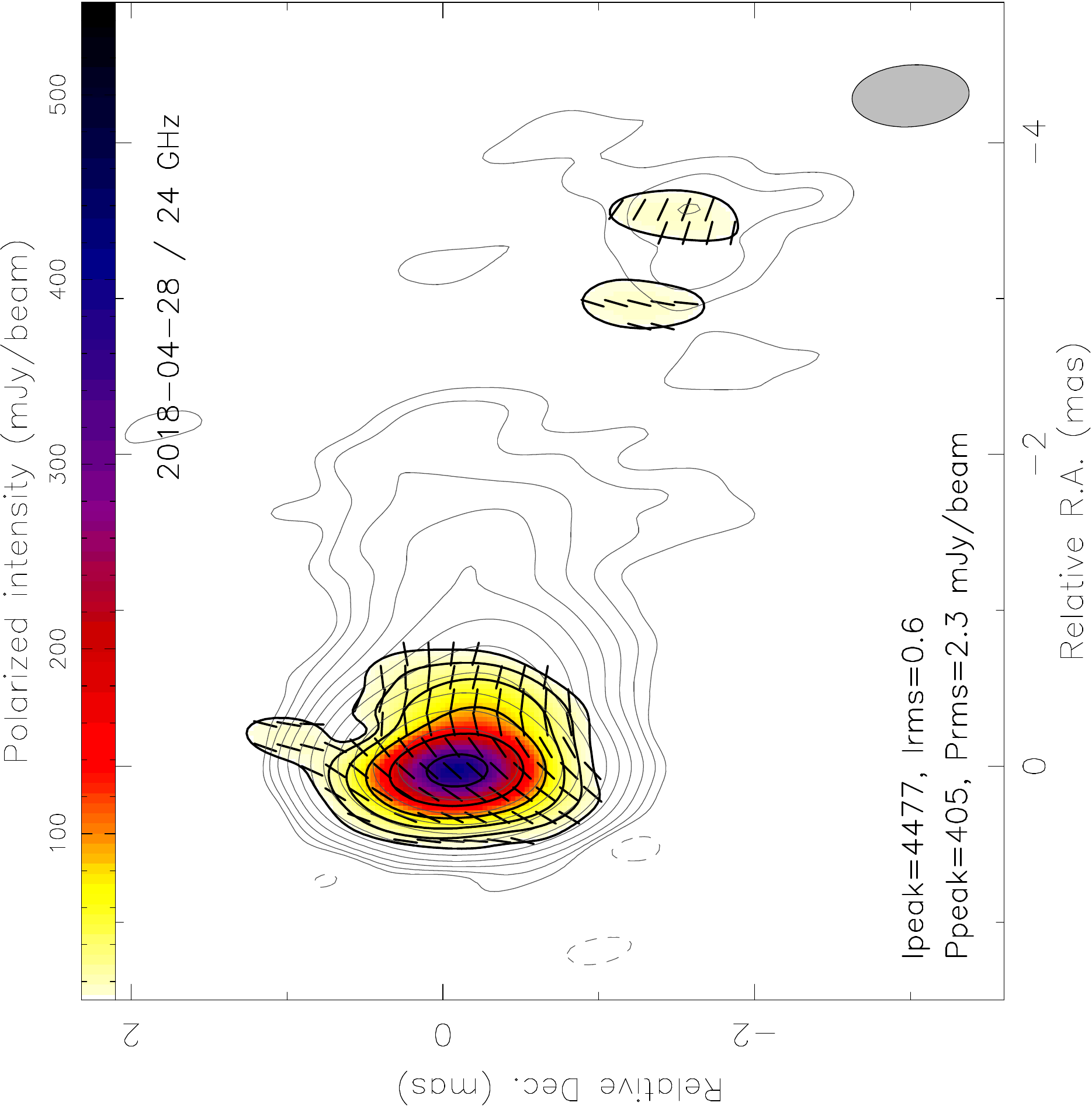}\\
   \includegraphics[width=0.54\columnwidth,angle=-90]{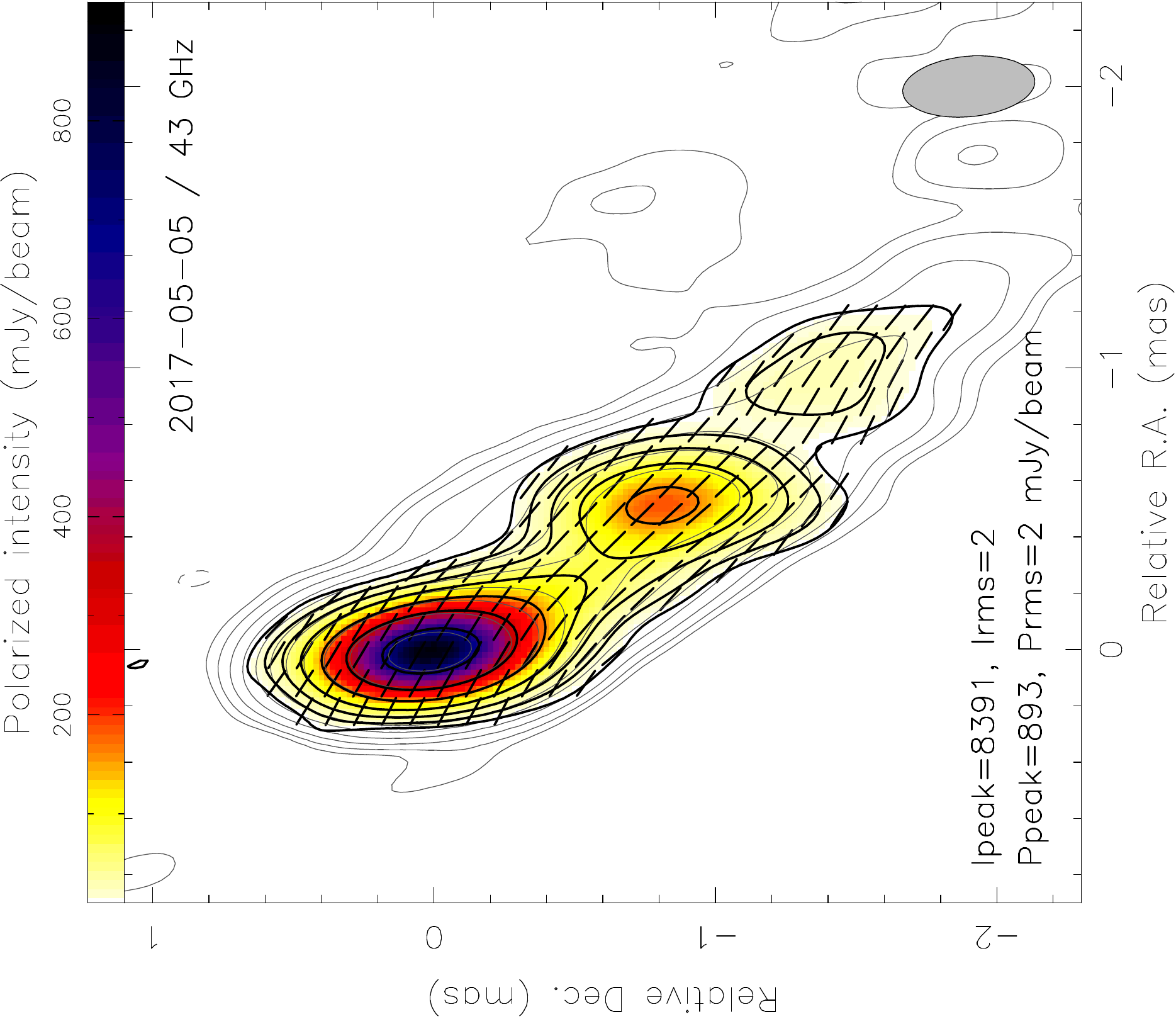}
   \includegraphics[width=0.54\columnwidth,angle=-90]{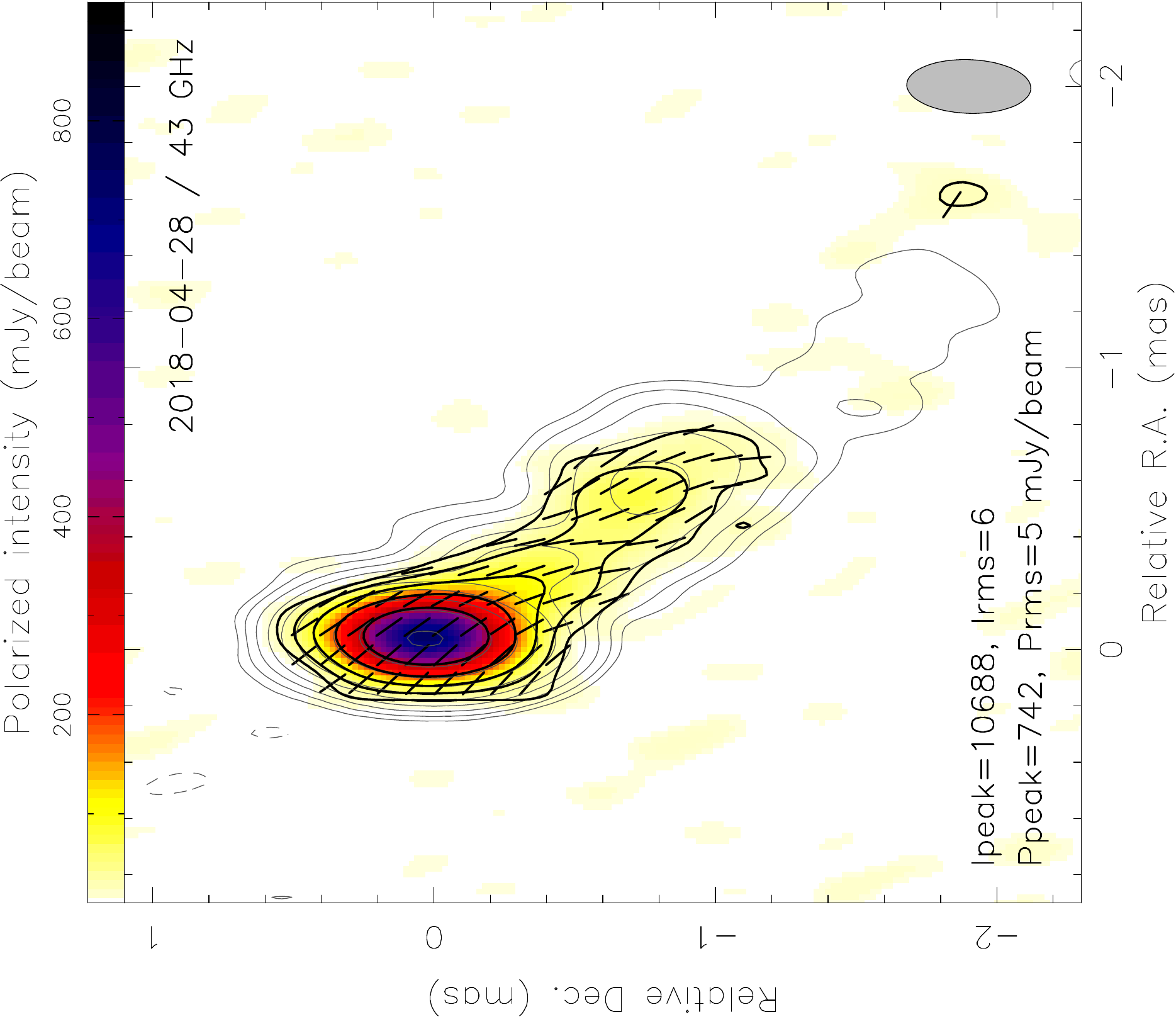}
   \includegraphics[width=0.54\columnwidth,angle=-90]{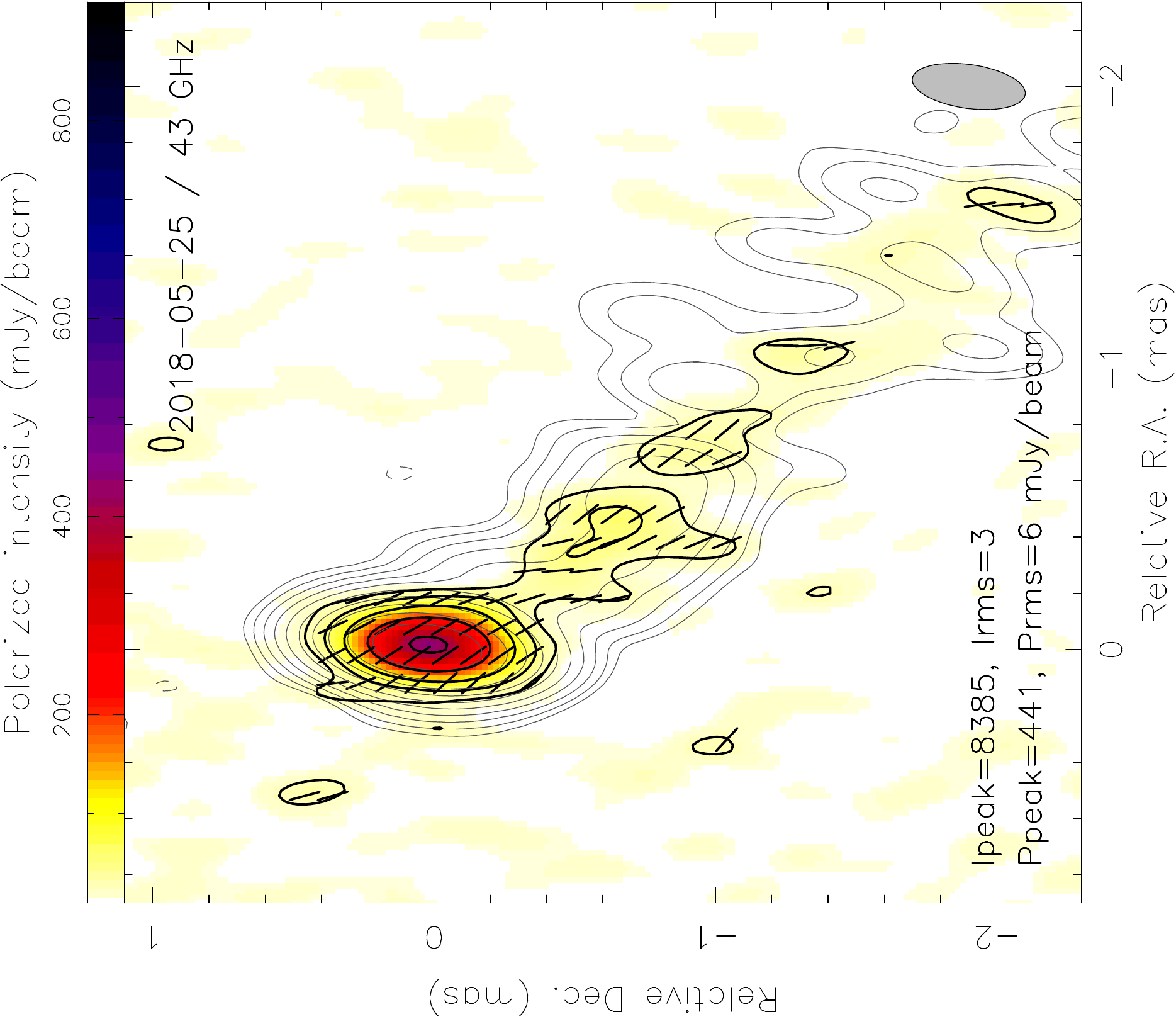}
   \includegraphics[width=0.54\columnwidth,angle=-90]{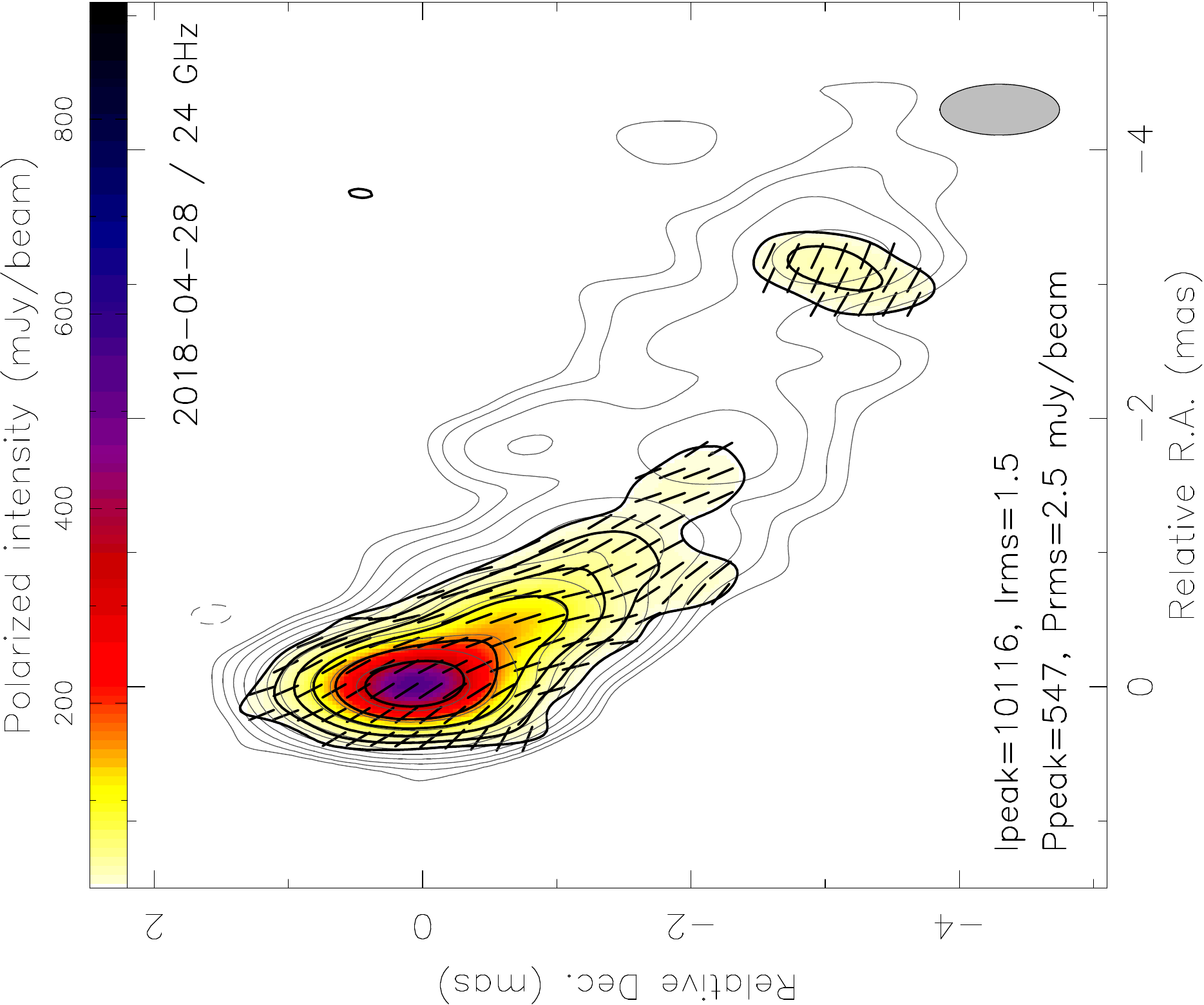}
   \caption{ Stokes $I$ images at 24 and 43\,GHz of OJ287 ({\it top row}) and 3C279 ({\it bottom row}), using D-terms determined from 3C279. }
   \label{fig:polcal}
\end{figure*}

Polarization leakage of the antennas (D-terms) was determined using the AIPS task LPCAL and a total intensity model of 3C279 (B1253$-$055), as using the Stokes\,$I$ model of OJ287 (B0851$+$202) gives less accurate estimates. 
We attribute this to a number of factors.
The most crucial one is related to the poor parallactic angle coverage for OJ287, which at some stations reaches 3\degr{} (see Table~\ref{tab:pac}).
Another factor is high fractional polarization, which at low signal-to-noise ratios  results in higher spurious polarization compared to a calibration with low fractional polarization \citep[e.g.,][]{2017AJ....154...54H}. 
In 2017, the fractional polarization of 3C279 ($\sim11$\%) was higher than that of OJ287 ($\sim7$\%). At the same time, 3C279 was brighter during this period, providing a higher signal-to-noise ratio compared to OJ287.
After 2017, the fractional polarization of OJ287 started to increase ($\sim11$\%), which is the opposite of the behavior of 3C279 ($\sim5$\%).
As a result, 3C279 provides better D-term estimates.
Finally, the polarized substructure is slightly more complex in 3C279 due to diffuse extended emission.
However, as seen in Fig.~\ref{fig:polcal}, EVPA is nearly constant across the sources during the dates of observations in both sources, making them comparable in this regard.
As a result, applying the D-term solutions from 3C279, we obtained broader dynamic ranges in the polarization compared to those from OJ287.

\begin{figure}
   \centering
   \includegraphics[width=0.7\columnwidth]{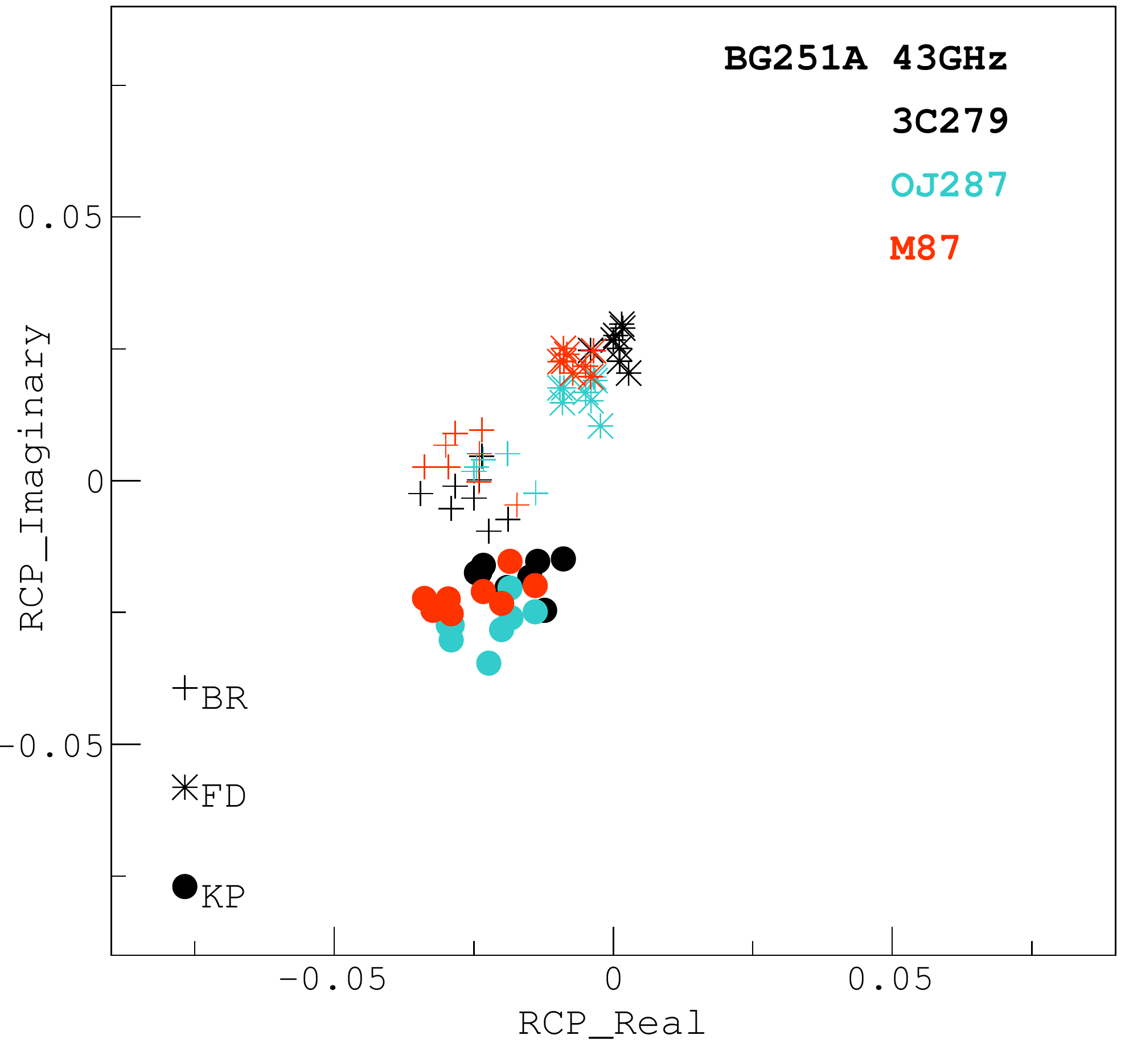}\\
   \includegraphics[width=0.7\columnwidth]{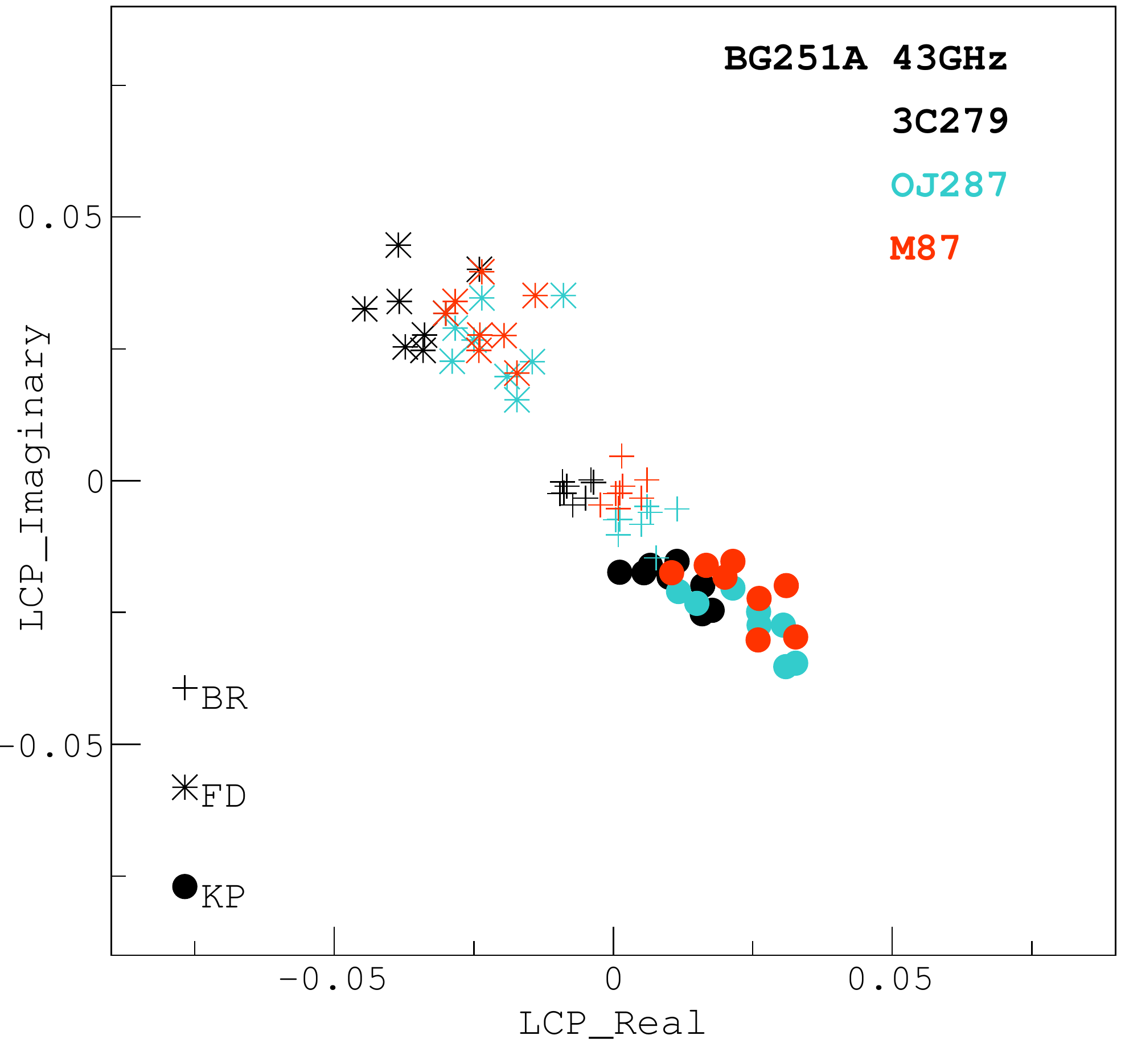}
   \caption{D-term measurements obtained for OJ287, M87, and 3C279 in 7\,mm VLBA observations on 2017 May 5 for RCP ({\it top}) and LCP ({\it bottom}) feeds at BR, FD, and KP. }
   \label{fig:dterms}
\end{figure}

The individual IFs were kept separate throughout the processing since the polarization leakage terms at each antenna differ among them.
The IFs were averaged together after calibration to make the final Stokes images.
We suggest that the higher sensitivity of polarimetric images obtained for the 2007 epochs (BW088A and BW088G) is due to frequency-dependence of the instrumental polarization within individual IFs. 
This problem became more serious with the IF bandwidth increase provided by the significant upgrade of VLBA hardware.

We determined the reliability of the final D-terms by comparing solutions obtained using OJ287, M87, and 3C279.
An example of D-terms for the BG251A experiment is shown in Fig.~\ref{fig:dterms}, while estimates for all epochs are given in Table~\ref{tab:dterms}.
The resultant parallactic angle coverage for OJ287, M87, and 3C279 is given in Table~\ref{tab:pac}.
D-term uncertainties, $\Delta\sigma_{\rm D}$, are estimated from the scatter between solutions determined from these three sources for each antenna and each feed (LCP and RCP), and resulted in an average value of 2\% at 24\,GHz and 1.5\% at 43\,GHz. 

We calculated errors in linear polarization using the  equations
\begin{equation}
\sigma_{\rm P} = \sqrt{\frac{Q^2\sigma_{\rm Q}^2 + U^2\sigma_{\rm U}^2}{Q^2+U^2}} \approx \frac{\sigma_{\rm Q} + \sigma_{\rm U}}{2},
\label{eq:sp}
\end{equation}
\begin{equation}
\sigma_{\rm EVPA} = \frac{\sqrt{Q^2\sigma_{\rm Q}^2 + U^2\sigma_{\rm U}^2}}{2(Q^2+U^2)} = \frac{\sigma_{\rm P}}{2P},
\label{eq:sevpa}
\end{equation}
where $\sigma_{\rm Q}$ and $\sigma_{\rm U}$ are the RMS values in the Stokes $Q$ and $U$ images, and $\sigma_{\rm P}$ and $\sigma_{\rm EVPA}$ are the uncertainties on the polarized intensity and electric vector position angle.
The feed calibration error is estimated following \citet{roberts_etal94} and \citet{2012AJ....144..105H} as
\begin{equation}
\sigma_{\rm D} = \frac{\Delta\sigma_{\rm D}} {(N_{\rm ant} N_{\rm IF} N_{\rm scan})^{1/2}} \sqrt{I^2 + (0.3I_{\rm peak})^2},
\label{eq:dterms}
\end{equation}
where $N_{\rm ant}$ is the number of antennas, $N_{\rm IF}$ is the number of IFs, $N_{\rm scan}$ is the number of scans with independent parallactic angles, and $I_{\rm peak}$ is the peak total intensity.
For our dataset the number of antennas varies from seven to ten depending on the epoch (see Table~\ref{tab:t1}), the number of IFs is eight, and the number of independent scans is six for 3C279 and seven for OJ287.
Errors on the polarized flux density and EVPA are therefore calculated using Equations~\ref{eq:sp} and \ref{eq:sevpa} by adding in quadrature to $\sigma_{\rm Q}$ and $\sigma_{\rm U}$ D-term errors defined by Equation~\ref{eq:dterms}.

The absolute orientation of the EVPA was determined using the EVPAs of OJ287 and 3C279 listed in the MOJAVE VLBA program\footnote{\url{http://www.physics.purdue.edu/astro/MOJAVE/index.html}} at 2~cm and the VLBA-BU-BLAZAR program at 7~mm \citep{2017ApJ...846...98J}.
Polarization observations at 15 and 43\,GHz were interpolated to produce position angles at 24\,GHz.
The uncertainty of the EVPA values provided by the monitoring program at 15\,GHz is 3\degr{} \citep{2012AJ....144..105H}, and at 43\,GHz is of 7\degr{} \citep{2005AJ....130.1418J}.
We therefore estimate our absolute EVPA values to be accurate within 7\degr{} at 43\,GHz and 8\degr{} at 24\,GHz.

\begin{table*}
\caption{Parallactic angle coverage for OJ287, M87, and 3C279 for 24 and 43\,GHz VLBA experiments.}
\label{tab:pac}    
\centering            
\begin{tabular}{llcccccccccc}
\hline\hline                 
Project& Source & BR& FD& HN& KP & LA & MK& NL & OV & PT &SC\\  
\hline 
\multicolumn{12}{c}{24\,GHz}\\
BG250AK & OJ287 & 81 & 123 & 31&  124& 108 &170  &69 &112 &--&17 \\       
        & 3C279 & 64 & 93 &  38& 93 & 84 & 82 & 66&84 &--&50 \\
        & M87   & 86 & 124 & 50 & 121 & 113 & 147 &99 &110 &--&154 \\       
\multicolumn{12}{c}{43\,GHz}\\
BG251A  & OJ287 & 83 & 123 & 32 & 125 & 108 & 13 & 70 & 112 & 110 & 3\\
        & 3C279 & 65 & 94  & 38 & 94  & 85  & 93 & 46 & 85  & 88  & 50\\
        & M87   & 86 & 124 & 91 & 121 & 113 & 147 & 99 & 110& 116 & 131\\
\hline         
BG250A  & OJ287 & 83 & 123 & -- & 125 & 108 & 170 & 70 & 112& -- & 18 \\
        & 3C279 & 64 & 93  & -- & 93  & 84  & 81  & 66 & 84 & -- & 50 \\
        & M87   & 86 & 124 & -- & 121 & 113 & 147 & 99 & 109& -- &154 \\   
\hline         
BG250B1 & OJ287 & 83 & 123 & 32 & -- & 95  & 178 & 70 & -- & 115& 17\\    
        & 3C279 & 64 & 93  & 38 & -- & 85  & 91  & 46 &--  & 88 & 51\\  
        & M87   & 86 & 124 & 91 & -- & 113 & 147 & 99 & -- & 116&154\\  
\hline
\end{tabular}
\end{table*}

\begin{table*}
\caption{Summary of the antenna D-terms (in \%) at 24 and 43\,GHz. Station abbreviations: BR = Brewester, FD = Ford Davies, HN = Hancock, KP = Kitt Peak, LA = Los Alamos, MK = Mauna Kea, NL = North Liberty, OV = Owens Valley, PT = Pie Town, and SC = St. Croix.}
\label{tab:dterms}    
\centering            
\begin{tabular}{llcccccccccc}
\hline\hline                 
Project& Feed & BR& FD& HN& KP & LA & MK& NL & OV & PT &SC\\  
\hline 
\multicolumn{12}{c}{24\,GHz}\\
BG250AK & RCP & 3.8$\pm$1.2 & 3.1$\pm$1.6 & 9.6$\pm$6.1 & 5.1$\pm$1.7 & 2.3$\pm$1.5 & 1.0$\pm$3.1 & 6.8$\pm$2.4 & 4.2$\pm$1.5 & -- & 5.0$\pm$4.3\\
        & LCP & 2.7$\pm$1.9 & 4.4$\pm$1.9 & 5.3$\pm$4.3 & 2.7$\pm$1.6 & 2.0$\pm$1.8 & 2.1$\pm$1.3 & 5.5$\pm$2.1 & 2.6$\pm$1.7 & -- & 5.4$\pm$6.3\\
\multicolumn{12}{c}{43\,GHz}\\
BG251A & RCP & 2.4$\pm$0.5 & 2.5$\pm$0.6 & 3.8$\pm$2.7 & 2.5$\pm$0.6 & 5.7$\pm$0.4 & 2.3$\pm$1.8 & 0.8$\pm$1.3 & 0.3$\pm$0.5 & 0.8$\pm$0.7 & 2.3$\pm$0.7\\
       & LCP & 0.6$\pm$0.7 & 4.6$\pm$0.8 & 2.0$\pm$1.0 & 2.3$\pm$0.9 & 4.1$\pm$0.9 & 1.2$\pm$0.4 & 1.1$\pm$0.6 & 1.3$\pm$0.7 & 1.2$\pm$0.9 & 3.1$\pm$2.4\\
BG250A & RCP & 1.5$\pm$1.1 & 3.5$\pm$1.8 & -- & 0.5$\pm$1.6 & 3.6$\pm$1.6 & 1.9$\pm$3.5 & 2.3$\pm$1.7 & 1.5$\pm$1.3 & -- & 3.6$\pm$7.0\\
       & LCP & 2.6$\pm$1.7 & 5.2$\pm$1.5 & -- & 0.5$\pm$1.7 & 2.8$\pm$1.8 & 1.0$\pm$2.1 & 1.4$\pm$2.1 & 2.9$\pm$1.5 & -- & 7.4$\pm$10\\   
BG250B1& RCP & 2.2$\pm$0.5 & 2.7$\pm$1.3 & 2.5$\pm$1.6 & -- & 5.0$\pm$0.6 & 1.7$\pm$2.1 & 1.9$\pm$1.8 & -- & 2.0$\pm$0.7 & 3.3$\pm$2.9\\
       & LCP & 1.5$\pm$1.2 & 4.6$\pm$1.1 & 1.7$\pm$1.7 & -- & 3.6$\pm$1.3 & 0.6$\pm$2.1 & 0.8$\pm$1.2 & -- & 2.7$\pm$1.0 & 5.9$\pm$4.1\\
\hline
\end{tabular}
\end{table*}

\section{Image alignment and spectral index}
\label{app:spi}

\begin{figure}
   \centering
   \includegraphics[width=0.6\columnwidth,angle=-90]{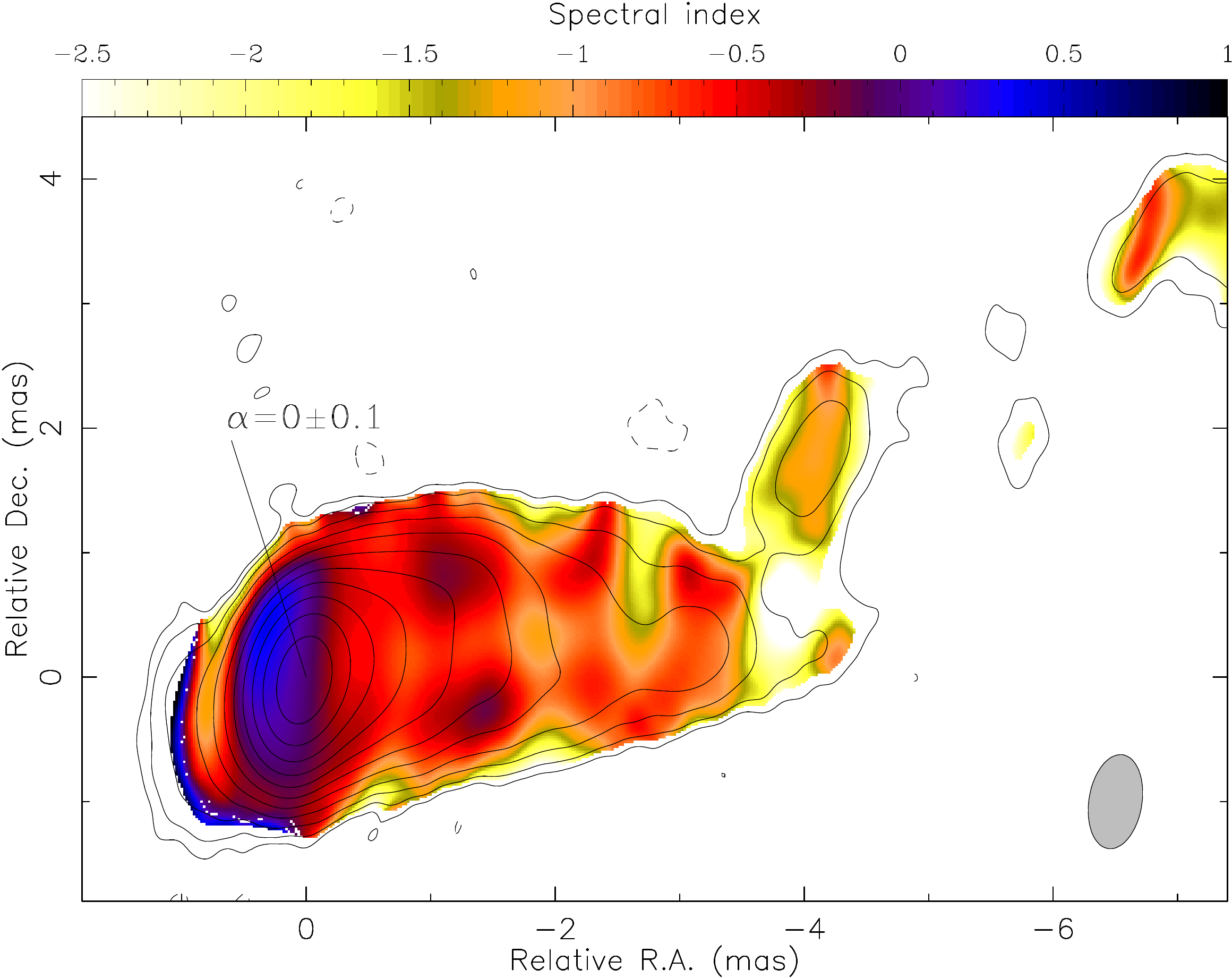}
   \caption{Spectral index (color), defined as $I\thicksim\nu^{\alpha}$, between the 24\,GHz and 43\,GHz total intensity images on 2018 April 28. Contours show the total intensity image from the 43\,GHz observations convolved with the 24\,GHz beam size.}
   \label{f:spi}
\end{figure}

Since the source has a complex structure, alignment of the images at two frequencies was made using the two-dimensional cross-correlation algorithm \citep{walker_etal2000}, which considers the whole optically thin emission region on the total intensity image.
The spectral index, $\alpha$, is computed such that $I(\nu)\thicksim\nu^{\alpha}$, where $I$ is the total flux density at observing frequency $\nu$.
The resultant image is shown in Fig.~\ref{f:spi}, and a value for the core region amounts to $\alpha=0.0\pm0.1$.
Considering the previous millimeter wavelength VLBI studies \citep{2009ApJ...697.1164B, 2015ApJ...807..150A, 2016ApJ...817..131H, 2018AA...616A.188K}, this implies that the synchrotron emission in the nuclear region of M87 is partially self-absorbed at 22 and 43\,GHz.
This is consistent with a nonzero shift of the M87 jet core with frequency in the 5--43\,GHz range \citep{2011Natur.477..185H}.

\section{Faraday rotation}
\label{app:frm}

\begin{figure}
   \centering
   \includegraphics[width=0.67\columnwidth,angle=-90]{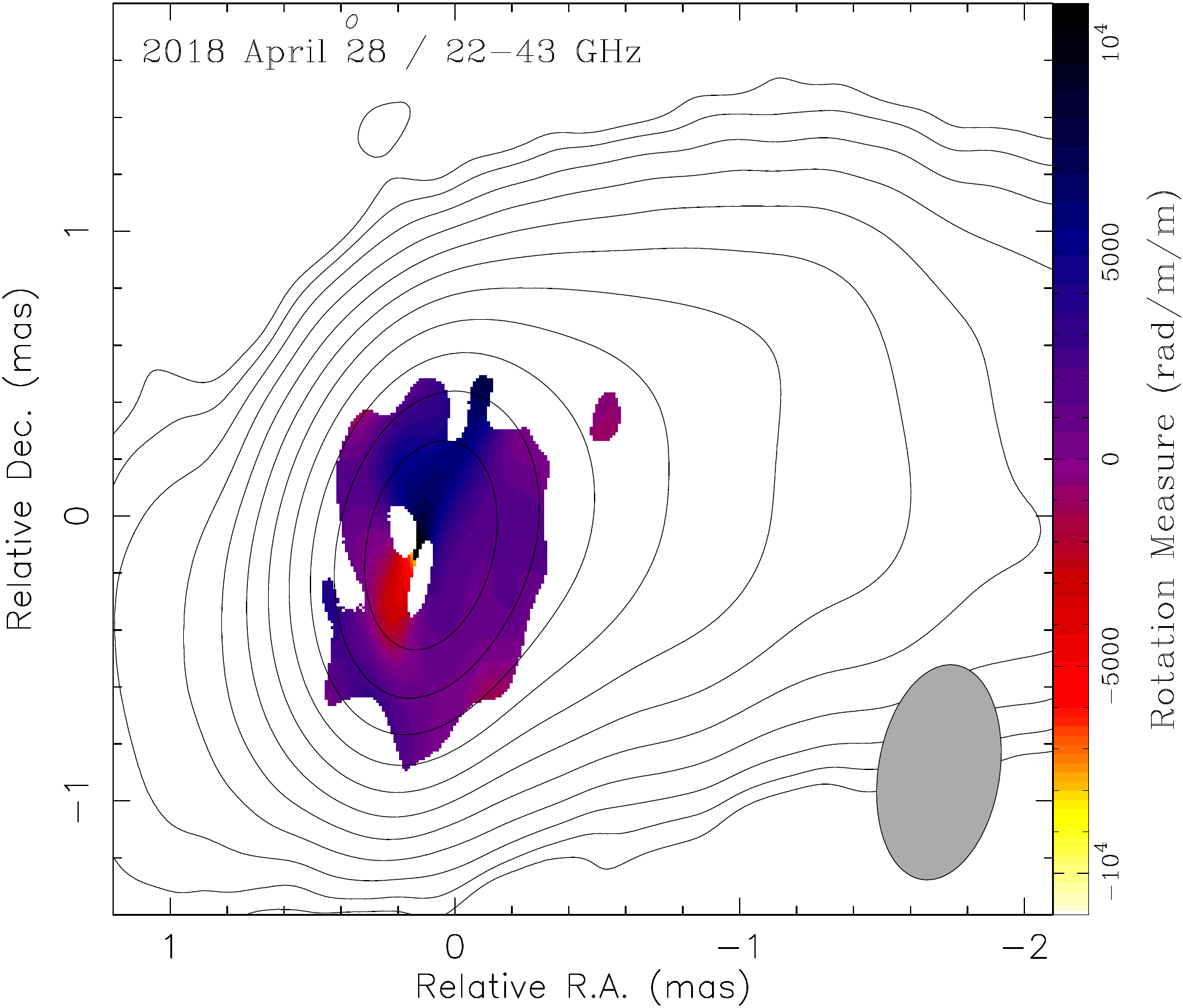}\\
   \caption{Rotation measure map using the 24\,GHz and 43\,GHz images. Color shows the RM in rad\,m$^{-2}$, and the restoring beam corresponds to that at 24\,GHz. 43\,GHz Stokes $I$ contours start from 0.38\,mJy/beam and increase by factors of 2, peaking at 712.88\,mJy/beam.}
   \label{f:app-frm}
\end{figure}

Figure~\ref{f:app-frm} shows the RM image which was computed under the
assumption of the smallest Faraday rotation, and relies only on the analysis of data at 24 and 43\,GHz.

\section{n$\pi$-ambiguity}
\label{app:pir}

Before determination of the slope of EVPA versus $\lambda^2$,   the $n\pi$-ambiguity in the angle needs to be solved. 
To do this we associate the linearly polarized feature of \citet{2016ApJ...817..131H}, located at $\thicksim0.1$\, mas downstream from the core, with the same region in our dataset.
This is a reasonable assumption because (a) polarized emission peaks at this region in our data through all epochs of observations at 24 and 43\,GHz, (b) the observed EVPA in this region shows only moderate variability over a long period ($\Delta\phi\sim20$\degr{} during 11\,yr), and (c) displacement of the 43\,GHz and 86\,GHz radio cores is insignificant \citep{2011Natur.477..185H}.
The best linear fit to the $\phi-\lambda^2$ dependence is therefore obtained when the 24\,GHz polarization angle is rotated by $-$180\degr{}, as shown in Fig.~\ref{f:frm}.
This corresponds to a reduced $\chi^2=3.44$ before and $\chi^2=0.55$ after the rotation is applied.

As an independent check of this $-$180\degr{} correction, we consider polarization measurements of M87 performed with the SMA at 230\,GHz on 2013 January 23 \citep{2014ApJ...783L..33K}.
They suggested that the observed polarized intensity originates in the innermost M87 jet, which is measured at a level of $\sim$1\% and oriented at $\sim30$\degr{}.
Using four individual intermediate frequencies in a bandwidth of 14\,GHz, \citet{2014ApJ...783L..33K} determined $RM=-(2.1\pm1.8)\times10^5$\,rad\,m$^{-2}$.
If we correct their measurements by $+180$\degr{}, and calculate RM between 86\,GHz and 230\,GHz, we obtain $RM=-(1.7\pm0.3)\times10^5$\,rad\,m$^{-2}$, which is in good agreement with the results of \citet{2014ApJ...783L..33K}.
Moreover, our solution for the $n\pi$-ambiguity in the angles provides overall better alignment of EVPA in the entire 24 -- 230\,GHz range of frequencies, as seen in Fig.~\ref{f:frm}.

\section{EVPA 90\degr{} -- flip due to opacity}
\label{app:opac}

\begin{figure}
   \centering
   \includegraphics[width=0.65\columnwidth,angle=-90]{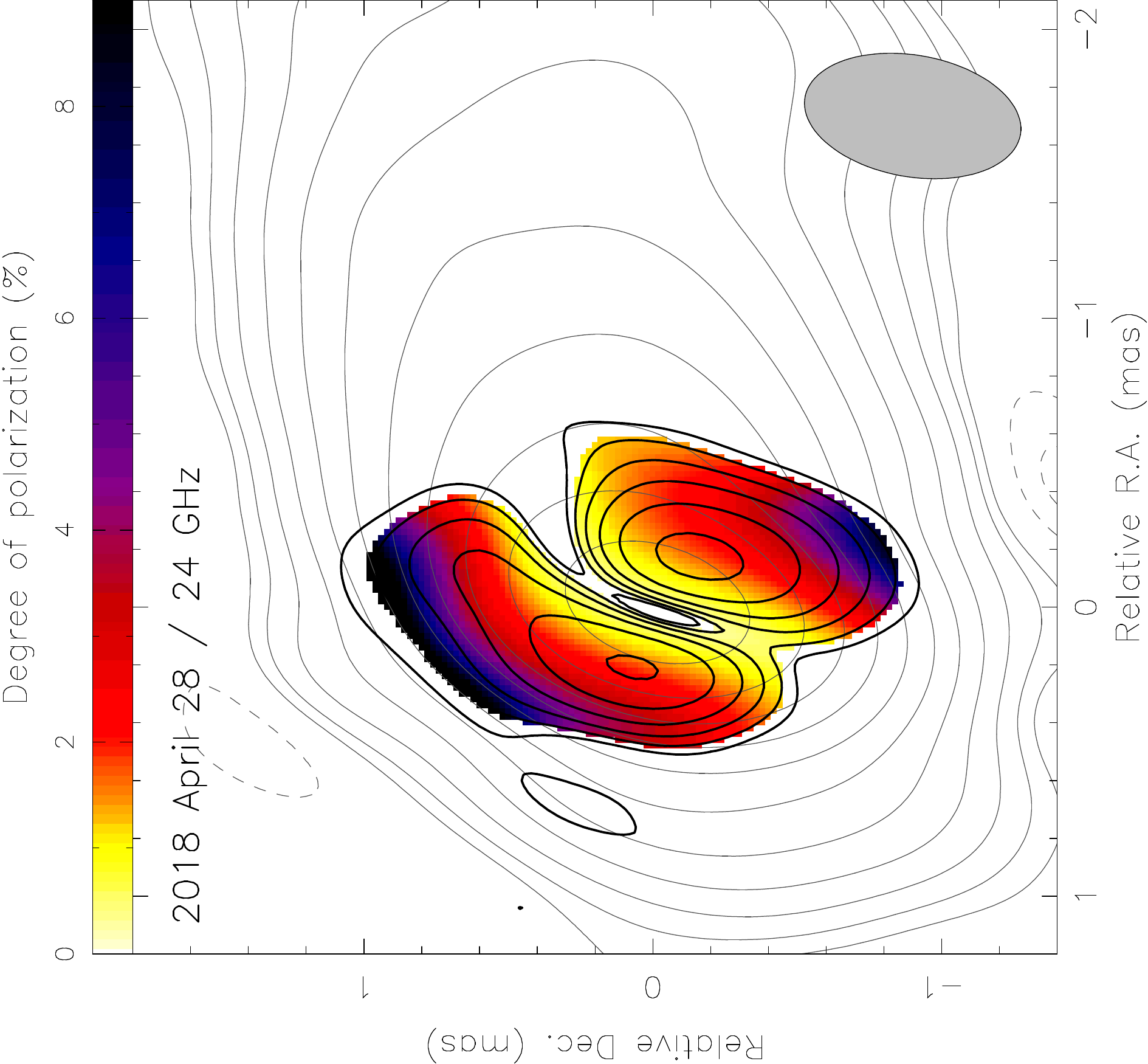}\\
   \includegraphics[width=0.65\columnwidth,angle=-90]{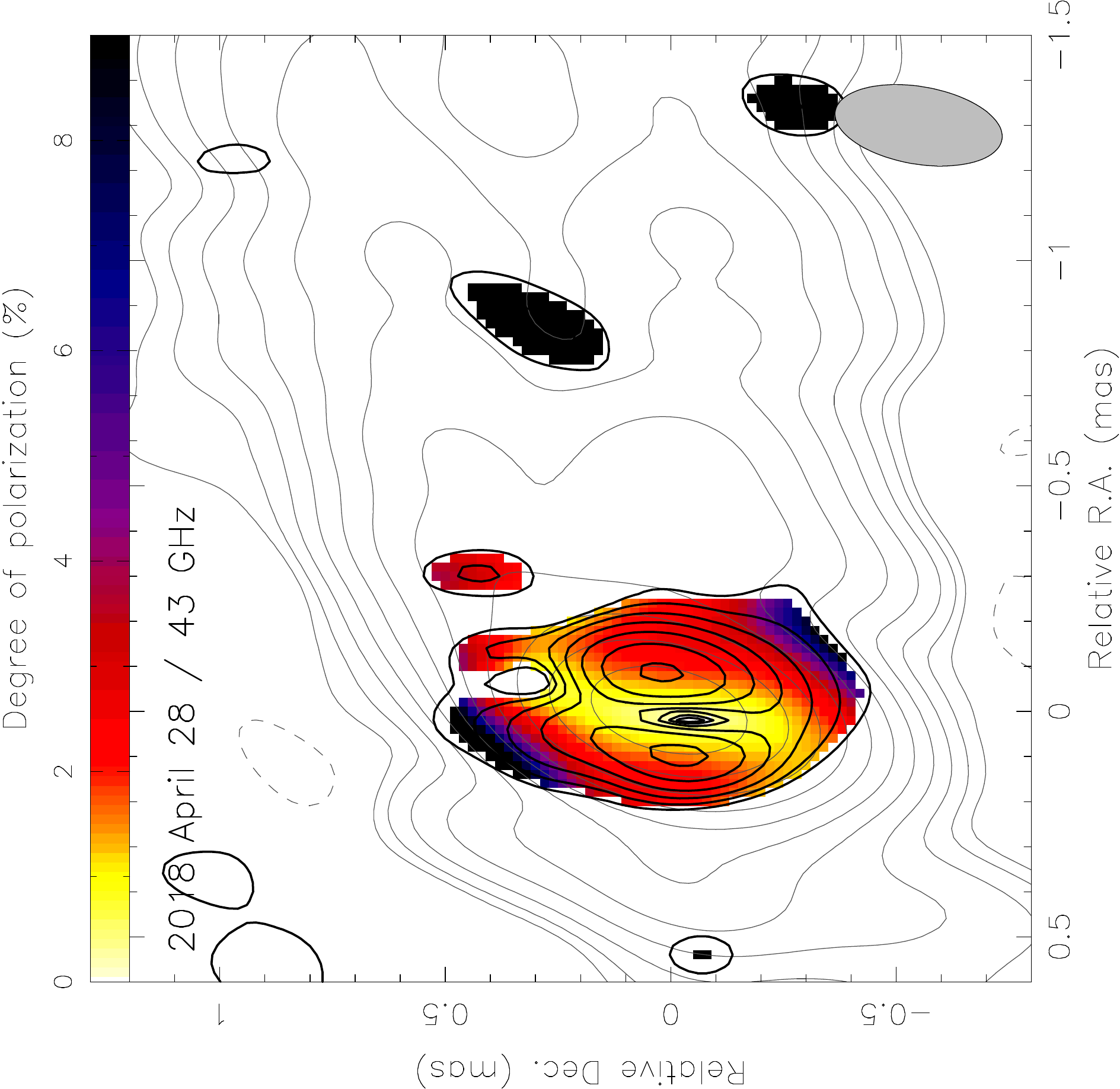}
   \caption{Degree of linear polarization (color) at 24\,GHz ({\it top}) and 43\,GHz ({\it bottom}) on 2018 April 28. Contours are the same as in Fig.~\ref{f:pol}.}
   \label{f:mdegr}
\end{figure}

It is commonly assumed that the radio core represents the surface of an optical depth $\tau\sim1$ \citep{1998AA...330...79L}, when synchrotron self-absorbed emission becomes visible.
Change in the emission regime from optically thin to optically thick is expected to be accompanied by a 90\degr{} flip of the electric vector, when the optical depth is large enough \citep[$\tau\sim6-7$,][]{1970ranp.book.....P, 2018Galax...6....5W}.
Due to the weak dependence of $\tau$ on the spectral index, it is unclear what the real value of $\tau$ is in this region of the M87 jet at 24--43\,GHz. 
Moreover, it is unlikely that optically thick emission from regions with such high opacities is ever observable in AGN jets, due to strong suppression of the emission.
According to \citet{1970ranp.book.....P}, the degree of polarization should significantly decrease (by a factor of $\sim7$) in the opaque regions, which is not observed (Fig.~\ref{f:mdegr}).
For these reasons, we omit the 90\degr{} correction.

\end{appendix}

\end{document}